\documentclass[
 aip,
 amsmath,
 amssymb,
 reprint,
]{revtex4-1}

\usepackage{graphicx}
\usepackage{dcolumn}
\usepackage{bm}

\usepackage[utf8]{inputenc}
\usepackage[T1]{fontenc}
\usepackage{mathptmx}
\usepackage{etoolbox}

\usepackage{enumitem}
\usepackage{caption}
\usepackage{color}
\usepackage{xcolor}
\usepackage{soul}
\usepackage{verbatim}

\usepackage{xurl}
\usepackage{hyperref}
\hypersetup{
    colorlinks=true,
    urlcolor=blue
}

\usepackage{float}

\usepackage[most]{tcolorbox}

\makeatletter
\def\@email#1#2{%
 \endgroup
 \patchcmd{\titleblock@produce}
  {\frontmatter@RRAPformat}
  {\frontmatter@RRAPformat{\produce@RRAP{*#1\href{mailto:#2}{#2}}}\frontmatter@RRAPformat}
  {}{}
}

\makeatother

\usepackage{nameref,hyperref}
\usepackage{lastpage,fancyhdr,graphicx}

\usepackage[most]{tcolorbox} 

\begin{document}

\preprint{AIP/123-QED}

\title{Systematizing cellular complexity: A Hilbertian approach to biological problems}

\author{Nima Dehghani}
\affiliation{McGovern Institute for Brain Research, MIT, Cambridge, MA, USA.}
\affiliation{Allen Discovery Center, Department of Biology, Tufts, MA, USA.}
\email{nima.dehghani@mit.edu}

\affiliation{$^{\dagger}$Current Address: McGovern Institute for Brain Research, Cambridge, MA, United States.}

\date{\today}

\begin{abstract}

Examining individual components of cellular systems has been successful in uncovering molecular reactions and interactions. However, the challenge lies in integrating these components into a comprehensive system-scale map. This difficulty arises due to factors such as missing links (unknown variables), overlooked nonlinearities in high-dimensional parameter space, downplayed natural noisiness and stochasticity, and a lack of focus on causal influence and temporal dynamics. Composite static and phenomenological descriptions, while appearing complicated, lack the essence of what makes the biological systems truly ``complex''. 

The formalization of system-level problems is therefore important in constructing a meta-theory of biology. Addressing fundamental aspects of cellular regulation, adaptability, and noise management is vital for understanding the robustness and functionality of biological systems. These aspects encapsulate the challenges that cells face in maintaining stability, responding to environmental changes, and harnessing noise for functionality. This work examines these key problems that cells must solve, serving as a template for such formalization and as a step towards the axiomatization of biological investigations. Through a detailed exploration of cellular mechanisms, particularly homeostatic configuration, ion channels and harnessing noise, this paper aims to illustrate complex concepts and theories in a tangible context, providing a bridge between abstract theoretical frameworks and concrete biological phenomena.
\end{abstract}

\maketitle

\small
\tableofcontents

\clearpage
\normalsize

\section*{Introduction}
\subsection*{The Complexity of Cellular Systems}
Unraveling the intricate web of processes in cellular systems presents a formidable frontier in contemporary biological research. Many significant breakthroughs in modern biology have come from studies at a molecular level, from genes to proteins. As our measurement techniques are providing us data at different scales with unprecedented resolution and throughput, a more detailed and nuanced picture is necessary for further advances, especially in areas where simplified models fall short of capturing the emergent properties and interactions within biological systems. Scientific journals and textbooks often feature graphical representations of cellular mechanisms, ranging from 3D depictions of subcellular structures to diagrams of signaling pathways. Advances in imaging and high throughput gene expression methods, coupled with a deeper understanding of biochemical pathways, have resulted in an abundance of intricate yet simplified representations of intracellular structures and signaling. However, it’s critical to recognize that these images are abstractions, simplifying complex cellular systems.

Capturing the complexity of cellular interactions with static descriptions is challenging. Given the complexity and abundant stochasticity in intracellular events, the ability of cells and simple organisms to achieve reliability and robustness is remarkable. In biological systems, many genes are responsible for coding sensors, actuators, and the intricate regulatory networks that manage them, providing robustness to variations rather than just the basic functionality required for survival in ideal circumstances \cite{Carlson2002}. These robustness mechanisms involve sophisticated regulatory feedback and dynamic processes that prevent cascading failures, creating systems so robust they appear simple, reliable, and consistently stable, seemingly unaffected by the environment \cite{Carlson2002}. Evolution favors the development of modular architectures with complex hierarchies of protocols and layers of feedback regulation, driven by the need for robustness in uncertain environments. These characteristics stem from the intricate relationship between complexity, robustness, modularity, and feedback \cite{Csete2002}.

The robustness of biological systems, which allows them to maintain functionality in the face of internal and external perturbations, has been extensively studied and is a fundamental characteristic of living organisms \cite{Kitano2004}. Robustness in biochemical networks, such as those governing gene expression, ensures that cells can function reliably despite the inherent noise and variability in these processes \cite{Barkai1997}. This robustness is often a result of complex regulatory networks that provide multiple layers of control and feedback mechanisms, allowing organisms to adapt to changing environments and maintain homeostasis \cite{Balazsi2011, Kitano2007}.

Understanding how cells control their inner dynamics and process environmental information, and how they respond to challenging conditions, may require more than reductionist or ateleological approaches. Referring to Lazebnik’s essay "can a biologist fix a radio" \cite{lazebnik2002}, we question whether we can truly understand the inner workings of a cell through conventional methods.

\begin{tcolorbox}[breakable, title=\textbf{Glossary}]
\begin{itemize}
    \item \textbf{Hilbertian Approach}: A method inspired by David Hilbert's list of 23 unsolved problems in mathematics presented at the Paris conference in 1900 \cite{Hilbert1902}. Hilbert's list was not merely a set of mathematical puzzles but a strategic roadmap to guide future research. This approach involves systematically framing key problems to direct theoretical development and research efforts. It emphasizes identifying fundamental questions and establishing a structured framework for scientific inquiry. In biology, as we propose in this work, this method can help in tackling the complexity and variability inherent in living systems by focusing on key unresolved questions that drive the field forward.
    \item \textbf{Axiomatizing Biology}: Axiomatization is the process of defining a set of basic principles or axioms from which other truths can be logically derived. In biology, this means establishing foundational principles that can explain various biological phenomena. Joseph Woodger and Mario Bunge made notable attempts to axiomatize biological sciences, though they faced challenges due to the complexity and variability of biological systems \cite{Mahner1997,Woodger1962}. 
    \item \textbf{Pearl's Framework}: A theoretical framework developed by Judea Pearl for causal inference, based on structural models and counterfactual reasoning. This framework helps in understanding causality in complex systems, providing tools to model and analyze causal relationships in biological networks \cite{Pearl2000,Pearl2009}.
    \item \textbf{Microscale, Macroscale, and Causal Levels}: Micro/Macro-scale terms refer to phenomena or structures at different scales of observation. Macroscale pertains to large-scale structures/phenomena, whereas microscale involves smaller, often microscopic structures/phenomena. To understand systems with dynamics across multiple scales, we turn to multiscale modeling. This approach combines models operating at different resolutions. Macroscale models provide a broader view, while microscale models offer intricate details. The goal is to strike a balance between accuracy and efficiency (macroscale models may not be accurate enough, while the microscale models may be inefficient.) \cite{Weinan2011}. Due to layered interaction across scales in biological systems \cite{Noble2008, Noble2012}, successful multiscale modeling hinges on causality. We must grasp how changes at one level influence behavior at another and develop quantitative models that can capture such `causal' interaction \cite{Hoel2013,Noble2008,Noble2012,Hoel2020,Hoel2017}. By connecting micro and macro scales causally, we create meaningful representations of complex biological dynamics.
    \item \textbf{Quantization Level}: In non-biological contexts, quantization involves representing analog signals as finite-resolution digital signals by slicing their amplitude into discrete levels. The reason for it is that the uniform quantization may not be optimal, as it rounds each sample value to the nearest value from a finite set of possible quantization levels \cite{Crecraft2002, Morris2011}. Biological systems, much like electronic circuits, encounter the need to represent continuous analog signals using discrete levels. These analog signals undergo quantization to facilitate processing, transmission, and analysis. Quantization involves dividing the amplitude range of an analog signal into distinct levels. The number of quantization levels determines the precision of representation. This can help achieving an optimal balance between precision and resource efficiency remains a challenge in biological contexts.
    \item \textbf{Teleology}: The study of purpose or design in natural phenomena. In biology, it refers to the explanation of biological processes in terms of their goals or functions rather than solely by their mechanistic causes. This concept, rooted in Aristotle's philosophy, has been critical for understanding evolutionary adaptations and the functional organization of biological systems \cite{Rosenblueth1943, Ayala1970}. During the Scientific Revolution, figures like Francis Bacon and René Descartes emphasized efficient causation over final causation, leading to a decline in teleological explanations \cite{McDonough2020}. However, William Harvey and Robert Boyle argued that teleological reasoning could complement mechanistic explanations, particularly in understanding biological functions. In contemporary biology, the debate continues, with some scholars framing properties of biological systems as \emph{``functions''} shaped by evolutionary pressures rather than purposeful design \cite{Wouters2003, Wimsatt19721}. Recent critique of selected effects theories, argues that biological functions should be seen as causes rather than effects, emphasizing the causal roles that functions play in biological systems \cite{Garcia2024}. It was the \emph{Cybernetics movement} that reintroduced teleological ideas, emphasizing feedback and control systems to explain goal-directed behaviors in biological organisms \cite{Rosenblueth1943, Wiener2019}. 
\end{itemize}
\end{tcolorbox}

\subsection*{Towards an Axiomatic Approach in Biology}
Many significant breakthroughs in understanding diseases at a molecular level have come from integrating insights across various scales, from genes to proteins to cellular networks. Thus, a more detailed and nuanced picture is necessary for further advances, especially in areas where simplified models fall short of capturing the emergent properties and interactions within biological systems.

In a quantitative approach to understanding the inner workings of cells, adopting methods from quantitative fields like physics or computer science can indeed be beneficial. However, we propose that a structured axiomatic framework can further enhance these efforts by providing a systematic way to address complex biological questions. This involves correctly framing questions and developing new methods to understand ``evolved'' biological systems. Integrating approaches from dynamical systems theory, computer science, statistical inference, and machine learning within this axiomatic framework can provide a more comprehensive understanding \cite{Vittadello2022}. Instead of concentrating on isolated issues or being confined by our expertise in certain methods, it is important to identify a spectrum of problems that a cell must address.

To address these challenges, this work proposes adopting a ``\emph{Hilbertian approach}'', where key problems are systematically framed to guide biological research and theoretical development. This involves identifying a spectrum of problems that a cell must address and developing new methods to understand these evolved biological systems. Developing a systematic theoretical approach not only is more productive but also shapes our thinking within a well-defined paradigm. This approach mirrors the influence of Hilbert’s list of unsolved problems in mathematics on the field, guiding our thought process in a similar manner \cite{Hilbert1902}.

Previous attempts at axiomatizing biology, such as those by Joseph Woodger \cite{Woodger1962} and Mahner and Bunge’s metaphysically-schemed approach to biology \cite{Mahner1997}, are prime examples of failed attempts due to improper formalization of key problems. To avoid repeating these mistakes, we would need to properly formalize these key issues before attempting to develop an axiomatic approach. By working within an scalable paradigm and adapting it as we progress (both experimentally and theoretically), we can bridge biological intuition with mathematical development, furthering our understanding of living systems. This approach can be very useful for avoiding the pitfalls of previous attempts and to ensure the successful axiomatization of biology. 

In light of these perspectives, we use cellular adaptability, ion channels, noise, and stochasticity as prototypical examples to illustrate the Hilbertian approach. This targeted exploration serves as a tangible context for applying and testing theoretical frameworks, bridging abstract concepts with concrete biological phenomena. By concentrating on these elements, we aim to construct a framework that can decipher the cellular machinery and predicts its behavior under diverse conditions. Under this carved path, the convergence of multidisciplinary methods could then enable us to transcend traditional barriers and unlock the enigmatic nature of cellular systems. Ultimately, we envision that attempts such as this endeavor and adopting a similar Hilbertian framework may pave the way for revolutionary advances in biology, offering insights that could transform our understanding of life at the molecular level through an axiomatic lens.

This paper aims to formalize key challenges in cellular biology: finding the right control switches, addressing the need for reconfiguration, and managing internal noise. By focusing on ion channels and their regulation, as well as exploring stochastic resonance, we provide an in-depth case study. Additionally, we outline other significant problems that could be part of a Hilbert-like set of questions in biology, advocating for a systematic and comprehensive approach to understand complex biological systems.

\section*{\label{sec1:level1}Problem 1: Where are the control switches?}
\subsection*{Challenges of Environmental Adaptation}
Single cell organisms must constantly adapt to changing environments, such as fluctuations in temperature, pH, and the concentration of necessary substrates or harmful chemicals. Multicellular systems have more internal stability due to sophisticated regulatory systems, but their individual cells must also overcome fluctuations in their constrained environments \cite{Selling2004,Skocelas2022}. A key question arises: How do cells, whether unicellular or multicellular, maintain their internal stability amidst these changes? 

Cells have developed intricate mechanisms to identify and manipulate relevant control mechanisms — or `\emph{switches}' — to adapt to environmental changes and preserve internal stability \cite{Selling2004}. Over time, in the context of evolutionary innovation, cells have developed regulatory motifs that allow them to efficiently manage internal and external changes. Network motifs provides valuable insights into these regulatory structures and their roles in cellular functions \cite{Alon2002,Alon2007,Alon2019}.
 
Effective regulation in cells involves navigating two critical constraints: firstly, the identification of these relevant switches without causing irreversible harm or death to the cell; secondly, the necessity of prompt action due to limited resources and time. Consequently, adaptive cellular regulation represents a time-constrained optimization challenge, centered around the evolutionary discovery and manipulation of the necessary `switches' for appropriate responses to environmental stimuli.

\subsection*{The Complexity of Cellular Switches}
Identifying the right switch at the right time poses a complex problem due to the vast array of potential switches in a cell. These switches can range from single molecular factors initiating signaling cascades to complex regulatory motifs and physical structures. A possible approach to this challenge is the concept of trial and error\footnote{Karl Popper highlighted the role of \emph{`trial and error'} as a fundamental process in knowledge acquisition within evolutionary epistemology \cite{Popper2005}.}, as posited by Ashby \cite{Ashby1945,Ashby2013}. He illustrated this with a hypothetical scenario in which an individual must find the correct configuration among 1000 switches to illuminate a light. The time required to find the right configuration varies significantly across different approaches. For instance, a thorough sequential search would take an impractical $2^{1000}$ seconds (assuming 2 seconds for each switch's on and off positions). In contrast, a serial test with some knowledge of partial correctness might average around 500 seconds, while testing all switches individually but in parallel could take merely 1 second. However, while this concept might be applicable to high-dimensional problems like adaptive cellular regulation, it comes with significant risks: incorrect switch activation, deactivation of essential switches, or time constraints might lead to detrimental outcomes for the cell.

In the context of cellular populations, one might speculate that some cells could successfully adapt through trial and error. However, the structure and hierarchy of cellular switches are far from random \cite{Yu2006,Tonner2015}. They exhibit a layered organization, with some operating at lower scales and others providing higher-level modulation. This hierarchical architecture implies a \emph{causal} modulatory tuning of more efficient, evolutionarily refined systems for achieving the right configuration, rather than relying on mere chance. This implies an evolutionary context where regulatory systems have been refined to ensure survival and adaptability.

\subsection*{Emergent Organization in Cellular Regulation}
In the context of cellular regulation, network motifs and complex switches play a pivotal role. Network motifs and complex switches provide functional building blocks within complex biological systems, occur at higher frequencies than in randomized networks, and are ubiquitous across various biological networks, including metabolic networks, gene regulatory networks, and in organisms ranging from plants to worms \cite{Defoort2018,Kreimer2008,Kashtan2005}. For example, motifs in gene regulatory networks of worms and plants have been found to function in a highly integrated manner, showing conservation and complexity across species \cite{Defoort2018}. Similarly, the evolution of modularity in bacterial metabolic networks demonstrates how modular structures can emerge to optimize metabolic efficiency and adaptability \cite{Kreimer2008}. These network motifs are not unique to cellular networks but also appear in ecological networks, indicating their fundamental role in maintaining stability and functionality in biological systems \cite{Stone2019}. Evolutionary processes have shaped these modular and motif-based structures to enhance the robustness and adaptability of biological systems \cite{Clune2013}. Network motifs can emerge from interconnections that favor stability, providing an evolutionary advantage in fluctuating environments \cite{Angulo2015}. Understanding these evolutionary processes that lead to the establishment of regulatory motifs can provide deeper insights into cellular regulation. Over time, cells have developed regulatory mechanisms that allow them to efficiently manage internal and external changes. Additionally, the identification of switches must be contextualized within the network of interactions in which they operate, as emphasized by the extensive work on network motifs \cite{Alon2002,Alon2007,Alon2019}.

Control mechanisms in biological systems often operate in a heterarchical rather than strictly hierarchical manner \cite{Bechtel2021}. This means that control is distributed across multiple levels and components, often with feedback loops and multiple controllers interacting without a single top-down directive. Simon's concept of nearly decomposable systems, where components at various levels interact in complex ways, also supports this view \cite{Simon1962}. This organized heterarchy highlights the flexible and adaptive nature of biological control mechanisms, which can operate independently yet coherently within the larger system \cite{McCulloch1945}. These complex control layers, combined with the dynamic and noisy nature of cellular environments, underscore that a simple trial and error approach is not efficient. Instead, the evolutionary refinement of modular and motif-based regulatory structures is crucial. This refinement ensures robust and adaptable responses to environmental challenges, allowing cells to efficiently manage internal and external changes. Given the inherent complexity of switch configurations, compounded by the often limited and noisy input information, makes a pure trial and error approach both computationally prohibitive and pragmatically risky \cite{Bei2013}.

\subsection*{Causal Inference in Cellular Regulation}
To reach a desirable problem-specific continuously evolving solution that requires little to no knowledge of the sources of external changes, cell would needs to be able to manage control levels at both macro and micro scales. This implies some form of internal tuning within the cell. Phenomenological models can describe the system at the macro scale, but they do not provide insights into the precise molecular interactions occurring at the micro level. Therefore, to understand the right switches at the right time is activated in a cell's exposure to environmental changes, it is necessary for us to conduct simultaneous measurements at various scales \footnote{A dichotomoy of view is shaping in the field; one in which, all these forms of internal control mechanisms are purely of evolutionary nature, and an opposing emergent view that adds a layer of learning and agential control to the single cell level control and decision making \cite{Lyon2021}. Interested readers may refer to a two-part special issue (in Philosophical Transaction B of Royal Society, 15 March 2021, Volume 376, Issue 1820) on this topic, entitled `Basal cognition: conceptual tools and the view from the single cell'.}. These scales should not be examined separately and then our static models of these scales be stitched together in a fixed diagram. Instead, overlapping complementary measurements at multiple scales can help construct a map of the causal structure of switches and reveal how the cell exerts internal control. This approach is more effective than creating an ever-growing chart of imaginary blueprints.

A methodological approach for uncovering the `causal' structure should consider the search for such motifs. Moreover, the reliance of regulatory information transfer on weak links, as opposed to the dominance of strong links in energy transfer and metabolic pathways, may have played a critical role in the evolution of new activating or inhibiting signals \cite{Kirschner1998}. The topological characteristics of these regulatory modules indicate varying connectivity structures, which are influenced by the degree of exposure to environmental variability and noise. Modules closely tied to environmental fluctuations tend to exhibit lower connectivity to enhance network robustness, whereas modules shielded from external noise sources lack a strong evolutionary impetus for sparse connections \cite{Navlakha2014}.

The complexity inherent in the multiscale nature of protein and genetic switches presents a formidable challenge in deciphering the causal mechanisms underpinning cellular control systems. In the realm of gene regulatory networks, this challenge is exacerbated by the limitations of current data types necessary for unraveling causal relationships \cite{Ahmed2020}. The ongoing evolution of methodological approaches, particularly in handling sparse and noisy gene expression time-series \cite{Aalto2020,Lu2021}, is crucial. However, the current state of data often leads to the development of methods that are inefficient or unreliable \cite{Ahmed2020}. As richer, higher-quality gene expression data becomes available, it becomes more important to focus on developing causal inference methods that are compatible with the complex, multiscale nature of cellular switch networks. For instance, in transcriptional regulatory systems, genes with low demand are predominantly regulated by repressors, whereas genes with high demand are controlled by activators \cite{Shinar2006}. This mechanism, aimed at minimizing error loads, might have been an evolutionary driver for the emergence of various regulatory motifs \cite{Shinar2006,Alon2007}. 

Considering this intricacy of interaction, search for causal links can be very insightful. Causal discovery methods have been applied to gene expression data to detect the presence and absence of causal relationships, even with very few samples \cite{Kang2010}. Similarly, methods developed for causal inference from gene perturbation experiments have validated the feasibility of discerning causal structures within gene regulatory networks \cite{Meinshausen2016}. Furthermore, single-cell analyses using tools demonstrate how causal discovery can elucidate gene regulation mechanisms at an unprecedented resolution \cite{Wen2023}. By prioritizing causal discovery within gene regulatory networks, we can highlight the underlying regulatory motifs that orchestrate cellular responses.

An exciting frontier in this field is the exploration of higher-order interactions and motifs. Higher-order interaction analyses reveal complex connectivity patterns that go beyond simple pairwise interactions \cite{Lotito2022}. These higher-order motifs can offer deeper insights into the modular organization of biological networks \cite{Battiston2021}. Data-driven approaches have also shown promise in identifying these intricate interactions within high-dimensional complex systems \cite{Tabar2024}. Combining these advanced techniques with the Pearl causal framework could push the boundaries of current biological research, uncovering novel regulatory mechanisms and evolutionary strategies.

Biological systems have evolved to minimize errors and maintain robustness amidst environmental and signaling noise \cite{Selling2004, Kitano2007}. This evolutionary resilience necessitates methods that can effectively capture causality from a functional hierarchical standpoint. A mere knowledge of connections between components is insufficient; instead, we require a form of causal probing akin to in-circuit testing \footnote{This approach is similar to how in-circuit testing for printed circuit boards works, where the testing is not just about understanding the connections, but also about probing the functionality of the individual components and the system as a whole. This allows for a more comprehensive understanding of the system, whether it’s a circuit board or a biological cell.} that can discern interactions at varying scales. Theories like Pearl’s framework of causal and counterfactual inference based on structural models \cite{Pearl2000,Pearl2009}, alongside its recent extensions incorporating information theory \cite{Hoel2013,Hoel2020}, can provide valuable tools for quantitatively analyzing causal structures. Recent advances have shown how macro-scales can emerge in biological systems and affect micro-scales, offering new avenues for prediction and control \cite{Hoel2020}. These tools can illuminate macro-scale models of network interactions.

However, the inherent time-delays and the complex ordering of processes resulting from genetic switch configurations pose significant challenges in unraveling the causal framework of intracellular switches. Presently, the mixed formalisms of information theory and causal modeling are not fully equipped to manage the daunting complexity of time-delays within extensive networks of genes and proteins. Concepts from concurrency theory, originally developed to address the problem of partially ordered components in distributed systems synchronization \cite{Lamport1978}, offer intriguing possibilities. Some concurrency-related methodologies, such as $\Pi$-Calculus and Petri nets, have been adapted to model cellular signaling pathways \cite{Regev2001,Priami2001,Pinney2003,Phillips2007,Priami2009}. Integrating these approaches with causal modeling and information theory could provide a more accurate representation of the nature of biological switches and their complex, intertwined multiscale dynamics. The primary focus should not be on what experts in a particular method can contribute to biology, but rather on first comprehensively understanding the biological problem and the cellular/subcellular constraints. Once these constraints are well-understood, we can then proceed to construct relevant and robust methods that are tailored to address these specific challenges. This approach ensures that the methods developed are not only scientifically sound but also practically applicable in the biological context.

\section*{\label{sec2:level1}Problem 2: How to manage the need to reconfiguration?}
\subsection*{Homeostatic response and feedback control}
In Problem 1, we explored how cells maintain internal stability through gene/protein switch configurations. Here, we delve deeper into the challenges cells face when managing significant extracellular environmental changes, necessitating not just maintenance but active reconfiguration of internal states. This problem focuses specifically on how cells reconfigure their regulatory mechanisms to maintain homeostasis under significant environmental changes, extending beyond the initial identification and manipulation of control switches discussed previously.
 
With minimal changes in the extracellular environment, a homeostatic response can handle internal regulation through a few adjustments. The concept of homeostasis, crucial for understanding this regulatory balance, was initially conceptualized by Claude Bernard and later formally defined by Walter Cannon \cite{Cooper2008, Cannon1929}. It deals with maintaining a steady-state goal as a core mechanism in living systems \cite{Modell2015}. This mechanism often relies on negative feedback \cite{Schneck1987,Modell2015}. The significance of negative feedback in achieving a dynamic equilibrium has been a pivotal aspect of cybernetics, highlighting how living systems autonomously counteract external changes \cite{Wiener2019}. In fact, cybernetics stretched this notion to define purposive behavior, giving the feedback regulatory mechanism a \emph{``teleologic''} flavor \cite{Rosenblueth1943}.

However, challenges such as lag time in response and dampening oscillatory offshoots prompted the development of other control mechanisms, notably integrative error minimization through negative feedback \cite{Yi2000}. Although mathematically elegant, the biological realization of integral feedback control was initially unclear. This gap has been recently addressed with the proposal of a controller topology enabling robust adaptation in noisy intracellular networks \cite{Briat2016,Aoki2019}.

\subsection*{Beyond Homeostasis: Adaptive Reconfiguration in Cells}  
Homeostatic regulation, whether through negative feedback or Proportional-Integral-Derivative (PID) control, is key in maintaining internal regulation in response to extracellular changes. Negative feedback helps achieve dynamic equilibrium by counteracting deviations from a set point \cite{Wiener2019, Rosenblueth1943}. PID control extends this by adjusting responses based on current state, history of past states, and rate of change, thus offering a robust mechanism for maintaining stability \cite{Chevalier2019}. Recent advancements have introduced a variation of this type of control, Antithetic Integral Feedback (AIF), which ensures robust perfect adaptation in noisy biomolecular networks by using two regulatory species that annihilate each other \cite{Briat2016, Aoki2019, Qian2022}. This approach has been demonstrated to achieve robust adaptation and has potential applications in maintaining homeostasis in complex intracellular environments. AIF controllers, in particular, may provide a universal solution for robust perfect adaptation, essential for coping with intrinsic and extrinsic noise in biological systems \cite{Aoki2019, Briat2016, Filo2022}.

However, the need for homeostatic regulation extends beyond handling moderate extracellular changes. In cases of more severe environmental shifts, such as a sudden but long-term change in extracellular Na, cells employ additional regulatory approaches that lead to transient expansion or contraction of the homeostatic response bounds \footnote{May be referred to as \emph{Allostasis} \cite{McEwen1998, McEwen1998b}, \emph{Heterostasis} \cite{Selye1973}, or \emph{Adaptive Homeostasis} \cite{Davies2016}.}. These adaptive responses, akin to integral control methods where the accumulation of error over time drives corrective measures \cite{Yi2000, Carlson2002}, are critical for cells facing acute or sustained environmental stresses, necessitating a reconfiguration of gene and protein pathways.

This not only pushes the cell to find the right switch configuration to maintain internal stability, but it also mandates the need for long-term or lasting reconfiguration. This leads us to conceptualize the cell’s response as a multidimensional landscape with regulatory stable states representing basins of attraction, akin to Waddington's landscape model \cite{Waddington1957} \footnote{
The Waddington landscape model offers a multidimensional perspective on cellular behavior, illustrating various potential states of cell development as attractor basins within a complex topography \cite{Waddington1957}. This model highlights the role of noise in shaping the epigenetic landscape, as fluctuations can distort the pathways that cells follow, leading to diverse cell-fate decisions \cite{Coomer2022}. Noise, therefore, is not merely a disruptive force but a crucial element that can drive variability and adaptability in biological systems. Modern interpretations of this model show how bifurcations and bistability are necessary for understanding the robustness and flexibility of cellular decisions and developmental pathways \cite{Ferrell2012}. The intrinsic noisiness of gene networks is another aspect where noise plays a pivotal role. The fidelity of molecular signaling is influenced by this inherent noise, impacting the information capacity and reliability of these regulatory systems \cite{Paulsson2004, Paulsson2005}. These perspectives collectively emphasize the importance of noise in both genetic regulation and epigenetic landscapes, illustrating how stochasticity can be harnessed for functional adaptability and complexity in cellular systems.}. Exposure to extreme events can push the cell to search for a new optimum basin in a new functional space with altered regulatory bounds, signifying a fundamental shift in the cell’s operational parameters.

Upon reconfiguration, a cell transitions into a new functional space, distinct from its original state. This new space presents the cell with unique and modified regulatory bounds, altering both the response time and nature of the intracellular response. The duration and severity of external changes dictate whether this reconfiguration is short-term or long-term. More complex cells may possess a broader functional repertoire, enabling them to navigate more basins in the functional space. This enhanced functional repertoire is underpinned by the biological robustness seen in higher organisms \cite{Skocelas2022, Selling2004}, which have evolved sophisticated regulatory networks and control mechanisms to maintain homeostasis and adapt to environmental changes \cite{Kitano2004,Aderem2005, Barkai1997}. For example, eukaryotic cells exhibit emergent macroscales in their protein interactomes, which are associated with lower noise and greater resilience compared to prokaryotic cells \cite{Barkai1997, Kitano2007} These higher-order structures allow for more effective information transmission and robust adaptability, indicating a significant evolutionary advantage in managing complex cellular processes \cite{Klein2020, Klein2021}. Therefore, the more advanced cells may have the capacity to quickly revisit an optimal basin in their functional repertoire upon re-exposure to a similar extreme event.

A key feature of adaptive homeostatic regulation is its predictive nature, which aids in reducing the magnitude of response error and the need for compensation \cite{Sterling2012}. In combination, homeostasis and adaptive homeostasis (allostasis) shape a functional space where phenotypic stability and plasticity define the dynamics and trajectory of a cell's configuration in response to varying degrees of extracellular changes.

\subsection*{Molecular Mechanics: \\ The Role of Ion Channels in Cellular Adaptation}
Ion channels serve as a prime example of the adaptive mechanism in cells. Essential for ion transport across cell membranes, these channels exhibit both phenotypic stability and plasticity, which are crucial for cellular adaptation. They control the membrane potential and intracellular ion concentration, both of which are dependent on the number and type of ion channels presently expressed in a cell. These channels are ubiquitous across unicellular and multicellular organisms, demonstrating the universal importance of this mechanism.

The diversity and complexity of ion channels reflect a higher functional repertoire in more complex cells, contributing to their ability to revisit optimal functional states quickly. This view can be scrutinized through an evolutionary lens. The evolution of ion channels highlights their importance in cellular responses to environmental changes. Early ion channels, such as calcium channels, evolved to include potassium and sodium ions and later fully evolved to dedicated potassium and sodium channels, which improved the precision and speed of signaling. This evolution was crucial for the development of efficient signaling mechanisms in multicellular organisms \cite{Franciolini1989}. Voltage-gated ion channels play a critical role in managing ion flow across cell membranes, which is essential for various cellular processes. These channels have diversified significantly, especially in more complex organisms, enhancing their ability to process and transmit information rapidly \cite{Moran2015}. Eukaryotes, in particular, have developed a vast array of ion channels that facilitate complex cellular behaviors. This expanded repertoire allows eukaryotic cells to manage a greater variety of functions with enhanced efficiency and adaptability, enabling them to respond more effectively to environmental changes \cite{Wan2021}.

In specialized excitable cells, such as neurons, the homeostatic regulation of ion channels is critical for maintaining excitability and compensating for perturbations such as channel deletions, mutations, or environmental disturbances \cite{Oleary2018, Williams2013}. These channels enable neurons to adjust their intrinsic properties through activity-dependent mechanisms that sense and respond to changes in physiological activity. They key point is that ion channels act as critical mediators between extracellular events and intracellular processes, coupling signals from the environment to cytoplasmic biochemical pathways. This capability underscores the advanced functional repertoire of more complex eukaryotic cells, allowing for refined and rapid responses to external stimuli \cite{Schauf1987, Anderson2017}.

To further illustrate these principles, let us consider the case of \emph{KscA}, a well-known bacterial potassium channel, is a tetramer with four-fold symmetry \cite{Doyle1998}. It operates as a pH-gated, potassium-specific channel, favoring a closed state at neutral pH. The channel exhibits voltage-gated inactivation, and the energy landscape of gating undergoes a conformational change, reaching its minimum when the channel is closed  \cite{Linder2013}. When the channel is exposed to an acidic environment, electrostatic changes in the protein’s cytosolic domain cause it to open \cite{Hirano2011,Baker2007}. The duration of the channel's inactivity influences its reactivation, thereby modulating its response to pH changes \cite{Gao2005}. 

Intracellular pH, regulated through changes in membrane permeability to ions like K+, Na+, and H+ \cite{Booth2007}, can affect many cellular mechanisms, including cell metabolism, cell growth, Ca2+ homeostasis, proliferation, and gene expression \cite{Putnam2012,Isfort1993}. 
Extracellular acidity impacts the cytosolic pH, altering the opening probability and activation of potassium channels such as \emph{KcsA}. This leads to a departure from the energetically favored closed state, placing an energy demand on the cell. The speed at which such a change of functional space occurs is controlled by the duration of prior inactivation.

In parallel to the response of voltage-gated K channels, other K transport systems play a crucial role in regulating the cytosolic pH, and thus its effect on intracellular K concentration. For instance, if a cell has Kdp, a K transport system sensitive to both extracellular K concentration and pH, its growth can be inhibited due to a joint reduction of Kdp activity and repression of Kdp gene expression caused by moderate external K+. Additionally, at lower pH, the cell's capacity to regulate its cytosolic pH through K transport is undermined.

This complex regulation involves intricate patterns of nested, hierarchical, and recursive control, where higher-level regulatory mechanisms modulate lower-level processes to achieve overall stability and adaptation \cite{Klein2020}. Nested and hierarchical structures provide a layered approach to control, while recursive functions enable dynamic adjustments based on feedback loops, enhancing the robustness and flexibility of cellular responses. The combination of these regulatory mechanisms renders the multiscale interactions inherent in cellular processes a complex entity for modeling and analysis.

The extensive literature bridging control theory and biophysics on ion channel dynamics, adaptation, and responses, provides valuable insights into the molecular constructs of these mechanisms. The control theoretic approach to modeling ion channel dynamics offers valuable insights into their function \footnote{Detailed reviews of mathematical models and simulations of ion channels can be found in \cite{Roux2004, Southern2008, Maffeo2012, Guardiani2022, Clerx2019}. These works explore various approaches, including traditional methods like Hodgkin-Huxley and Markov models, as well as modern techniques such as Artificial Neural Networks (ANNs) and Neural Ordinary Differential Equations (ODEs) for modeling ion channel conductance and kinetics \cite{Jeong2021, Lei2021}. These advanced models offer improved accuracy and computational performance, providing valuable tools for studying the electrophysiological mechanisms of ion channels across different biological systems.}. System theory-based models, for instance, provide a novel method for characterizing ion channel kinetics, allowing for exceptional accuracy and computational efficiency compared to traditional methods \cite{Langthaler2022}. The application of control theory concepts to neuronal homeostasis has clarified many underlying theoretical aspects and has suggested new experimental and computational approaches to better understand these processes \cite{Oleary2011}. While control theory and physics can guide us in untangling the dynamics at given scales, we also need formal systems that can aid in linking the models and can untangle the complex web of interactions in these forms of nested, hierarchical and recursive structures. As such, a recent work on category theory offers a formal mathematical framework to understand the compositional structures, providing tools for the detailed analysis of nested and hierarchical systems in cellular processes \cite{Dehghani2023}. Similarly, formal treatment of recursion in partially ordered components in distributed systems \cite{Lamport1978} can provide the needed complementary tool for deciphering complex regulations that was discussed above. By adapting and fusing these approaches, we may better understand the need for cellular reconfiguration in response to environmental changes.

\subsection*{Adaptive Plasticity in Complex Cellular Systems}
In more complex cells and multicellular organisms, the interplay between phenotypic stability and plasticity is further complicated by additional layers of regulatory control. For instance, the steady-state distribution of ion channels is governed by a correlated homeostatic regulation \cite{Oleary2013}. Furthermore, the macroscopic electrogenic function can be maintained despite variation in ion channel types \cite{Ori2018}. The co-regulation of ion channels through homeostatic mechanisms illustrates how different channels can compensate for each other to maintain stability. This co-regulation is essential for managing the complex interplay between multiple properties and ensuring robust \emph{macroscopic} cellular function. The emergence of correlations in ion channel expression levels can be explained by homeostatic control mechanisms that couple expression rates to cellular activity \cite{Yang2022, Oleary2013}. 

On the other hand, other complicated parameters can also set the functional landscape of cellular functioning. For example, in neurons, there can be differential expression of Na and K channels between soma and dendrites \cite{Dumenieu2017}. This suggests that the functional space and its minima are uniquely defined for soma versus dendrites. As a result, the functional space in dendrites or soma could be maintained within a pre-set range while it might be altered for the other parts of the neuron. This differential plasticity allows the neuron, as a unit of computation, to have varying overall function due to a polarized spatial balance between phenotypic stability and plasticity. 

Such differential regulation underscores the dynamic nature of cellular responses and the importance of understanding these responses in the context of the entire organism. Additionally, intrinsic membrane excitability can be altered through dynamic epigenetic modification of DNA (through methylation) that controls the expression of ion channels and synaptic response \cite{Meadows2015, Meadows2016}.

A similar situation is observed in cardiac tissue where the balance between phenotypic stability and plasticity is controlled by a variety of mechanisms at different levels \cite{Rosati2004}. 
For example, repetitive pacing transforms the electrophysiological attributes of cardiomyocytes to a new semi-stable state \cite{Meadows2015, Meadows2016}. This shift in the functional space occurs through a reduction in the mRNA concentration of INa and ICa channels, effectively altering their expression in cardiomyocytes \cite{Yue1999}. The complex molecular mechanism for electrical remodeling of the cardiac cells creates a major challenge for understanding how fine-tuning ion channel expression controls cardiac excitability \cite{Balse2012}.

The phenomenon of altering electrophysiological properties in excitable membrane (of neurons or cardiomyocytes) can be studied through mixed models combining recursion and dynamics. These models provide a deeper understanding of the intricate causal chains affecting a cell's functional space post reconfiguration. This approach underscores the need for robust, interdisciplinary methods to unravel the complexities of cellular adaptation and reconfiguration in response to environmental changes.

\section*{\label{sec3:level1}Problem 3: How to harness noise rather than succumb to it?}
\subsection*{The Dual Nature of Cellular Noise}
Cells face the challenge of noise at almost every scale of operation. Noise exists at the boundary with the external world as well as internally across molecular interactions and physical interfaces. Despite the abundance of noise, cells manage to function, grow and evolve. The viewpoint that noise is pure nuisance has been gradually shifting to an understanding that noise is not necessarily detrimental to biological systems \cite{Tsimring2014}. In this context, we will explore various aspects of noise, how cells handle such challenges, and what modeling frameworks can elucidate this complex relationship. While it may seem narrow to focus on specific aspects, this approach allows us to delve deeply into critical examples that illustrate broader principles. By examining specific instances of noise handling, we can uncover underlying mechanisms and strategies that are broadly applicable across different biological systems, thereby maintaining coherence with the holistic perspective advocated throughout the article.

The cell membrane acts as the boundary between the internal and external worlds of the cell. Through molecular sensing and transport, cells receive information about the external world and harvest necessary chemicals. Ligand-receptors and ion channels are key molecular systems through which a cell communicates and samples the external world. Aside from cases where a ligand irreversibly binds to its receptor, the ligand-receptor interaction is inherently stochastic and noisy \cite{Azpeitia2020}. This inherent noise, contrary to being a mere byproduct, plays a crucial role in the cell’s information processing capabilities. The irreversible binding may be desirable for drug testing purposes \cite{Newman1990}, but in principle, is the opposite of what a continuous information acquisition system would need. Considering the ligand-receptor binding as the means for probing the extracellular space, one can think of the ``source (of the chemical), particles, and receptor'' as an equivalent of ``transmitter, channel, receiver''. An information-theoretic treatment can guide in separating different noise sources in a mixture model of kinetic and stochastic reactions \cite{Pierobon2011}. 

In the context of negative feedback and genetic networks, a framework combining information theory and control theory provided insights on the fundamental limits on the intracelluar noise-reduction \cite{Lestas2010}. By applying similar principles, we can better understand how cells manage noise at the various steps of signal transduction. However, at each level, if not accounting for the subtleties of biological systems, a blind adaptation of such methodology might fail to capture many nuances. The speed of ligand-receptor binding defines the nature of the ``noise'' from a specific vantage point. Slow binding and unbinding can cause the response curve to gradually reach a saturation point, making it challenging for the cell to differentiate between high ligand concentrations \cite{Teimouri2020, Duke1999}. Receptor noise studies have demonstrated that slow binding leads to reduced chemotactic efficiency, establishing bounds on the fidelity of such signaling processes \cite{Rappel2008a, Rappel2008b}.

Furthermore, from the perspective of the cell, measuring high concentrations of ligand introduces a different kind of noise management compared to low concentrations. At high concentrations, receptors can become saturated, making it difficult to distinguish between slightly different high ligand levels. This saturation arises not from low molecular counts but from limitations in receptor availability and binding dynamics. Cells can utilize pre-equilibrium sensing and signaling, where information from early, transient binding events is used before the system reaches equilibrium \cite{Ventura2014}. This allows cells to bypass the saturation limit, enhancing discrimination between high ligand concentrations and improving signal fidelity. Conversely, when the number of receptors in a cell is low, another form of noise is introduced due to the low signal-to-noise ratio caused by the small number of ligand molecules. In such cases, the shift of response tuning to downstream machinery can prevent the amplification of receptor-level noise, maintaining accurate signaling \cite{Ventura2014}.

\subsection*{Boundary Noise and Cellular Communication}
Another special treatment of noise occurs at the boundary, where interference noise due to the presence of structurally-similar non-target molecules adds to the inherent stochasticity of a small number of target molecules. In such situations, ligand-induced stochastic cluster formation of receptors takes place, enhancing the receptor's sensitivity and specificity \cite{Kajita2020}. 

This ``noise-cancelling'' receptor clustering and ensued allosteric interactions have evolved to enhance the signal-to-noise ratio (SNR) in cellular communication. Clustering of receptors increases sensitivity by reducing the detrimental effects of intrinsic and extrinsic noise. For example, receptor clustering has been shown to increase signal detection reliability and transduction efficiency, thereby optimizing the cellular response to environmental cues \cite{Aquino2011, Care2013}. Additionally, ligand-induced stochastic cluster formation of receptors enhances both sensitivity and specificity through noise-induced symmetry breaking, compensating for the presence of non-target ligands in the environment \cite{Kajita2020}. In certain cases, this can be understood through the concept of stochastic resonance \footnote{Stochastic resonance, where noise actually enhances the detection of weak signals, is further elaborated in the following sections.}. Particularly in clusters of major histocompatibility molecules and T-cell receptors, a stochastic quantizing encoding of transmembrane signaling improves information transfer in cellular processes \cite{Bene2022}.

A solution for managing noisy ligand-receptor interactions is to synchronize the excitable receptors during signal transduction \cite{Nagano2020}. These noise-induced phenomena, whether they result in spatial symmetry breaking or temporal symmetry breaking, might have separate evolutionary roots. However, they hint at a key point: cells are not merely passive receivers faltering in the face of abundant noise at the boundary. An important lesson is to recognize that the nature of the noise itself can guide us in modeling and understanding it effectively. As Monod elegantly put it, biological systems are shaped by deterministic laws and randomness \cite{Monod1974}. While reductionist approaches provide detailed insights into the molecular aspects of noise in biological systems, they often fall short in capturing the emergent properties of complex systems. In biological contexts, the separation of scales is not straightforward. Higher-level properties and processes are not merely derivable from lower-level data and mechanisms but also act as causes of lower-level behavior, involving interactions in both directions \cite{Noble2021}. Understanding how cells sample information from their environment and manage their responses to noise requires considering multiple levels of organization and these causally linked scales. This integrated perspective helps to elucidate how cells harness noise and adapt to changing conditions, thereby maintaining functionality and adaptability in changing environments.

However, the way that living systems harness noise, as exemplified above, shows us that a purely reductionist approach to the biophysical characteristics of ligand/receptor/membrane will leave us ill-equipped in understanding how cells manage challenging environments \cite{Noble2021}. While reductionist approaches provide detailed insights into molecular interactions, they often miss the emergent properties and cross-scale dynamics that are crucial for understanding noise management. By integrating both reductionist details and perspectives on interscale interactions, we gain a more comprehensive understanding of cellular resilience and adaptability.

A deeper understanding of boundary noise and its management is necessary for understanding cellular efficiency in noisy environments. Developments such as patch clamp technology have revealed the conductance fluctuations of individual channels as they oscillate between open and closed states, influenced by thermal noise \cite{Hille2001,Sakmann1995}. The probability of voltage-gated channels opening and closing is dependent on the membrane potential, which in turn affects channel noise. These fluctuations impact the timing and generation of action potentials, as well as subsequent changes in membrane potential. Channel noise can cause significant variations in spiking propagation \cite{Horikawa1991}, and it imposes a lower limit on the miniaturization of cell signaling \cite{White2000,Schneidman1998}. This is because the noise generated by ion channels can shift the functional state of the cell, leading to spontaneous action potential generation \cite{Faisal2005}. Understanding these dynamics not only elucidates how cells maintain functionality under noisy conditions but also sets the stage for exploring how noise can be harnessed to enhance cellular processes, as discussed in the following section on stochastic resonance.

\subsection*{Harnessing Noise: Stochastic Resonance in Cellular Function}
While some aspects of noise can be limiting or detrimental, noise itself can enhance the detection of weak signals in certain nonlinear systems, such as electronic circuits and biological sensory systems \cite{Wiesenfeld1995}. A phenomenon known as \emph{stochastic resonance} (SR) has emerged as a key player in biological systems dealing with noise. SR, which enables the amplification of weak signals in nonlinear systems, illustrates that noise, when managed appropriately, can enhance cellular functionality \footnote{SR was first introduced as an explanation for the observed periodic occurrences of the Earth's ice ages \cite{Benzi1981}. In SR, adding noise to a nonlinear dynamical system can bring a weak signal above the threshold, enabling the system to detect sub-threshold signals \cite{Nicolis2007,Gammaitoni1998}. In this case, the system that is subject to periodic forcing shows a resonance in the spectrum that is absent in the forcing and the perturbation \cite{Benzi1981}.}. 

SR has been shown to contribute to signal detection in various sensory systems, including acoustic/electric stimulation of human hearing \cite{Zeng2000}, vision \cite{Simonotto1997}, and noise-enhanced tactile sensation and balance control \cite{Priplata2002,Collins1996}. Large nonlinear networks are more sensitive to weak inputs when a fixed level of noise is added. In such cases, irrespective of the nature of the input signal, it can be pushed to cross the threshold \cite{Collins1995}. 

However, SR is not limited to large networks of excitable units. SR has also been observed at the cellular level in crayfish mechanoreceptor cells \cite{Douglass1993}, and in bistable single neuron models that show a correlated switch between states when they are driven by noise in the presence of periodic external modulation \cite{Bulsara1991}. For SR to be effective at this scale, the noise must be optimized and adjusted to the nature of the signal \cite{Wiesenfeld1994}.

Stochastic resonance (SR) also plays a significant role in ion channel signal transduction at the subcellular level. This was first demonstrated in a large parallel ensemble of an artificial ion channel, specifically polypeptide alamethicin incorporated into planar lipid bilayers. In this case, the signal transduction induced by noise increased over a hundred-fold \cite{Bezrukov1995}. Ion channel conformational changes are considered as thermally-activated Poisson events. Ion channels open and close stochastically, and they do not have a built-in ``absolute" threshold. Ion channels open and close stochastically, and they do not have a built-in “absolute” threshold. However, additive noise linearly changes their activation barriers with applied transmembrane voltage, thereby increasing the instantaneous rate of the thermally activated reaction and exhibiting SR \cite{Bezrukov1998}. This phenomenon allows for the detection of small amplitude input signals in thermally driven physico-chemical systems, where reaction rates are controlled by activation barriers. This is applicable in systems such as semiconductor p--n junctions and voltage-dependent ion channels \cite{Bezrukov1997}.

Although the activation barriers in ion channels change linearly with additive noise, the downstream effects in these driven nonlinear systems demonstrate complex behavior. The presence of noise can induce resonant amplification of weak signals, a hallmark of SR, which significantly enhances signal detection and transmission. This is because SR exploits the interplay between noise and the nonlinear dynamics of the system, leading to emergent behaviors that are not predictable from the individual components alone \cite{Vincent2021}. This interplay often reveals new dynamic states and responses \cite{Sorokin2021, Lucarini2019}, highlighting the importance of considering higher-order interactions and emergent properties to fully understand noise management in biological systems.

Although, it's worth noting that the artificial channels used in these studies are less sensitive to temperature variations than natural channels \cite{Schmid2001,Parc2009}. The higher temperature sensitivity of natural channels near physiological temperature, and the induced conformational transitions, effectively change the energy landscape and transition rate between closed and open states at the level of single ion channels \cite{Parc2009}. As a result, at the scale of a single ion channel, enhanced thermal-noise-induced SR affects the rate of information gain more significantly in biological ion channels than in artificial channels \cite{Goychuk2000, Parc2009}.

Periodic, aperiodic, and nonstationary stochastic resonance occurring at physiological temperature allows cells to harness noise to enhance signal transmission and information gain with fewer channels \cite{Goychuk2000,Adair2003}. This strategy is energy-efficient, as increasing the density of ion channels to combat noise is energy-consuming.

The collective property of ion channels and its effect on SR also influences the response to noise. The number of ion channels determines the amplitude of the membrane potential fluctuation \cite{Toral2003}. In small clusters of ion channels, SR and threshold-crossing are primarily defined by individual channel kinetics and thermal-noise-induced SR \cite{Schmid2001}.

As the size of the membrane patch increases (for example, in soma vs dendrites) and/or the density of ion channels increases (through modulation of gene expression or their polarization, as in neurons), a system-size resonance occurs \cite{Pikovsky2002}. This emergent collective property of globally coupled ion channel assemblies exhibits a resonant-like temporal coherence \cite{Schmid2001}. This collective property leads to a type of SR that occurs only in large clusters of ion channels, even in the presence of the suboptimal intrinsic noise of single channels \cite{Schmid2001,Schmid2004}.

\subsection*{Quantization and Noise Management in Cellular Systems}
Unlike engineered systems, where efforts are focused on eliminating noise, cells harness boundary noise at the subcellular scale. SR plays a key role in dealing with boundary noise at both the individual ion channel and collective levels.
By examining the challenges that cells must overcome, we can gain a deeper understanding of the computational primitives that operate through cellular biophysical interfaces and biochemical pathways. The stochastic nature of ion channels and the presence of stochastic resonance can guide us to probe information gating through the lens of quantization error.

Ligand receptors and ion channels serve as quantizing sampling devices. Their stochastic kinetics and density define the collective sampling rate. We have discussed how SR can enhance the sensitivity of ion channels to subthreshold inputs. Equally important, thresholded systems exhibiting SR are unique forms of dithered quantizers \cite{Wannamaker2000}. Dithering \footnote{During World War II, engineers discovered that mechanical computers used for radars and trajectory calculations performed better onboard airplanes than on the ground. They found that vibration-induced noise randomized the quantization error, thereby increasing the accuracy of the calculations performed by these mechanical computers \cite{Pohlmann1989}. Since then, it has become standard practice to add a small amount of noise to a signal either before quantization (during analog-to-digital conversion) or when reducing the bit-depth of a digital signal (such as during a 256-bit to a 16-bit image reduction).}, a technique commonly used in digital signal processing domains such as audio and picture coding, involves adding random noise (or dither) to data before storage or transmission. This process randomizes the quantization error when a reduction in precision is necessary \cite{Roberts1962}. This dithering characteristic of SR allows cells to modulate the sampling precision and information transfer of ion channels through a quantization operation.

Thresholded dynamical systems, such as ion channels and excitable membranes, can be interpreted as multistable systems. Signal transduction through these channels has a finite range, and the sampling process is discrete and stochastic. This leads to inevitable distortion and loss of signal details, which can manifest as spurious frequencies or amplitude reduction due to time discretization and amplitude quantization.

In systems with a nonlinear input-output relationship, the quantization error needs to be independent of the input signal to minimize distortion \cite{Gammaitoni1995_a}. In thresholded nonlinear dynamical systems, like cells with excitable membranes, stochastic resonance (SR) is more akin to a special case of a dithering effect rather than a resonant phenomenon \cite{Wannamaker2000,Gammaitoni1995_a}.

Just as image dithering maintains the color-depth gradient during bit reduction (e.g., when converting grayscale to black-and-white), SR in multi-thresholded dynamical systems can increase the number of quantization levels \cite{Gammaitoni1995_b}. SR affects the responsiveness of individual channels, causing them to open stochastically at different rates. Collectively, these opening dynamics create a multi-threshold system.

Instead of setting a fixed threshold for all channels to switch to the open state, the presence of noise induces concentrated transitions to the open state in clusters of channels. The small intervals created around multiple thresholds, determined by the different sizes of these clusters, increase the number of quantization levels. This increased quantization level allows the cell to transduce microscopic fluctuations in the external environment in a graded, discretized manner.

As a result, by harnessing noise, cells can effectively capture a macroscopic view of their environment at the boundary. We can infer that SR enables cells to be sensitive to microscopic fluctuations while maintaining a macroscopic understanding of their environment.

\subsection*{Modeling and Multi-Level Implications of Cellular Noise}
Cells' ability to operate efficiently amidst noisy dynamics underscores the functional role of noise in biological systems \cite{Azpeitia2020,Tsimring2014}. The fact that noise at various levels can exhibit non-rigid functional characteristics is significant. This includes boundary stochastic noise in a single channel, clusters of channels in a membrane patch, collective properties of polarized distribution of channels in a single cell, and networks of cells. This highlights the importance of multi-level (multiscale) modeling.

Multiscale modeling and simulation provide powerful tools to capture the complexity of biological systems \cite{Weinan2011}. Developing a common framework for multiscale modeling, which includes defining clear scales and ensuring proper scale bridging, helps in understanding how microscopic interactions aggregate to influence macroscopic phenomena and vice versa \cite{Hoekstra2014}. A balanced approach that integrates reductionist and systemic perspectives is necessary for understanding complex systems \cite{Laughlin2000}. While reductionism provides detailed insights into individual components, emergent properties and interactions across scales will require a different approach to understand collective behaviors that are not evident from the properties of individual components. This integrated approach ensures that modeling techniques are tailored to specific contexts, recognizing the variability in scale interactions across different biological systems \cite{Rice2024}. Given the spatial variability and temporal dynamics of the involved components, we need to resort to the heterogeneous multiscale method in order to effectively couple macroscopic and microscopic models \cite{Weinan2003}. In cellular systems, this method can help capture how molecular-level interactions impact macroscopic behavior. For instance, deciphering the dynamics of ion channels and receptors at the microscopic level can significantly clarify our understanding of complex macroscopic phenomena like cell signaling, membrane excitability, and cell-cell communication in the presence of noise.

The multiscale effects of stochastic resonance align well with the concept that there is no privileged level of causality in biological systems \cite{Noble2018}. This suggests that all levels of biological organization, from molecular to cellular to systemic, play a crucial role in the overall function and behavior of the system. This perspective encourages a comprehensive approach to studying biological systems, taking into account the interplay between different levels and the role of noise in these interactions. By integrating insights from molecular dynamics, cellular interactions, and system-level behavior, we can better understand the emergent properties and complex interactions and how cells harness and manage noise to maintain functionality and adaptability. 

\section*{\label{sec4:level1}Concluding Remarks}
\subsection*{The Challenge of Biological Modeling}
Hilbert’s list of problems was not just a set of unsolved mathematical problems, but also a framework for advancing the discipline. While some of these problems remain unsolved, the concept of a framework has endured. Unlike mathematics, which is grounded in logic and proofs, biology has largely been an ever-expanding collection of detailed findings with limited scope. Efforts at biological modeling and theorizing have frequently struggled to synthesize these findings into unified theories. This raises the question: What is missing in our approach? Is biology fundamentally different from quantitative fields like physics? Can biology not be tamed with mathematical models? or is it the case that we yet do not have the appropriate mathematical tools to properly model biological systems?

In a thought-provoking essay \cite{Gunawardena2014}, Gunawardena posits that our models, at least at the molecular scale, are mostly phenomenological and amount to little more than educated guesswork. He points out that unlike physics, biological models are not objective. Therefore, our best strategy is to begin with a sound set of assumptions that can guide us in selecting the most suitable model for the problem at hand \cite{Gunawardena2014}. The essay paraphrases James Black, a Nobel laureate pharmacologist, reminding us that ``our models are accurate descriptions of our pathetic thinking''. This depiction of biological models highlights key issues. Biological modeling faces additional challenges due to the complexity and variability inherent in living systems. The dynamic interactions, nonlinearity, and stochasticity at multiple scales make it difficult to create comprehensive models that are both predictive and explanatory.

\subsection*{Embracing a Problem-Oriented Approach in Biology}
One of the challenging issues with current biological models is their focus on isolated components rather than integrated systems. This approach leads to disjointed models that hinder our ability to formulate fundamental theories. While biological sciences have seen exponential growth, the progress in finding principles has not paralleled that fast-paced progress. Some biological textbooks even include words like ``fundamentals'' or ``principles'' in their titles, yet they primarily expand with more detailed examples and new findings without consolidating overarching principles (As an example, the 1st edition (1981) of ``Principles of Neural Science'' was published in 468 pages; and its 6th edition (2021) has now grown to 1646 pages). This massive expansion in volume, while reflecting incredible progress in terms of our knowledge about some details of the system, is antithetical to extracting fundamental principles in theory-laden science. This phenomenon underscores a significant challenge: the need for models that bridge different scales and components, and the need for a framework that would facilitate integration across such models.

Here, we seek to address this challenge by proposing a program inspired by Hilbert’s approach. By grounding our understanding in fundamental principles and axioms, we can systematically build multiscale models that capture causal links across different scales. This problem-oriented approach keeps us aligned with empirical data and ensures the development of  \emph{falsifiable} models, fostering a cohesive framework for integrating diverse biological models into core principles. Adhering to constraints and details rooted in the physical nature of biological components is crucial. This adherence will give us a chance to formulate models that are true to the underlying biophysics with predictions that can be experimentally falsified \cite{Bialek2012}.

Necessarily, we need to avoid being trapped by disjointed descriptions of biological systems. A computational lens can guide us to focus on systemic functions rather than isolated components. There are convergent themes between biological and computing systems that fall under the lens of ``computational thinking,'' such as distributed and coordinated processes, robustness to failure/attacks, and modularity \cite{Navlakha2011}.

The concept of separation of scales, where slow processes are treated as static constraints, dynamic processes are governed by mechanical laws, and fast processes are averaged over, is crucial in modeling many physical systems. However, in biological systems, scales are often coupled and causally-linked, making the separation less straightforward. Complex systems in biology require an integrated approach that considers the interactions across scales. This integrated approach is helpful for capturing the true complexity of biological systems. By emphasizing problems over methods to guide proper heterogeneous multiscale models, we can identify the constraints inherent in these problems and develop integrated models that reflect the true complexity of biological systems. This approach aligns with the idea that biological theory can benefit from combining parsimonious reasoning with rule-based explanations \cite{Krakauer2011}. By incorporating these principles, we can better understand how to build effective models that bridge different scales and components, providing a more comprehensive understanding of biological systems.

\subsection*{Integrating Biophysics, Computational Thinking \\ and Cybernetic Principles}
The proposed framework, synergizing computational thinking with biophysical modeling, aims to foster the design of experiments towards uncovering fundamental biological principles. This approach also addresses the limitations of cybernetics by integrating insights on feedback and control within a broader reductionist perspective of molecular biology. While cybernetics correctly recognized the crucial importance of feedback and control in biological systems \cite{Wiener2019}, its overemphasis on teleology \cite{Rosenblueth1943} juxtaposed it with the explosive success of reductionist molecular biology. What makes biology truly unique is its concern with purpose \cite{Mayr2004}, but perhaps searching for behavioral purposeful descriptions untied from the underlying biophysical processes was a key behind cybernetics' demise. We can bring back the insightful aspects of cybernetics to the center stage and avoid its pitfalls.

It is important to acknowledge that models solely aiming to outperform their competition in data fitting are non-falsifiable \cite{Gunawardena2014}. The increasing usage of artificial neural networks (ANNs) in biology is noteworthy. While ANNs have shown success in modeling complex communication and control pathways, their success hinges on a high-parameter nonlinear fitting process. This often leads to the replacement of one complex nonlinear system with another black box that remains largely misunderstood. However, the potential of ANNs can be harnessed more effectively within the framework of mathematical models, where they can serve as nonlinear optimizers. This approach enables the fusion of physical laws with data-driven methods, yielding a more comprehensive description and prediction of system behavior \cite{Rackauckas2020, Karniadakis2021}. Importantly, it also ensures the falsifiability of our theories, a cornerstone of scientific modeling. 

Advancements in machine learning present opportunities to integrate data-driven methods with physical principles, facilitating the creation of models capable of deciphering complex dynamical systems from noisy and incomplete data \cite{Lu2022,Reinbold2021}. Combining machine learning with multiscale modeling can enhance our understanding of complex biological systems by leveraging the strengths of both approaches. Multiscale modeling provides a structured framework to address different temporal and spatial scales, effectively capturing the interactions between them \cite{Groen2019}. For instance, this approach involves identifying relevant scales and developing models at each scale, which are then combined to form a comprehensive multiscale model. This method allows us to understand how microscopic interactions influence macroscopic phenomena and vice versa \cite{Weinan2003}. Integrating machine learning with multiscale modeling can address the limitations of each approach individually. Machine learning excels in handling large datasets and uncovering correlations, while multiscale modeling is adept at probing causality and mechanisms \cite{Alber2019}. By combining these methods, we can create robust predictive models that integrate underlying physics to manage ill-posed problems and explore massive design spaces. This integration allows for a more comprehensive understanding of biological systems, capturing emergent properties and complex interactions across scales \cite{Vlachas2022}.

This integrated framework signifies a substantial opportunity for advancement in biology. Given the current technological landscape, encompassing precision devices, high-throughput techniques, and advancements in computing and AI, we are well-positioned for significant progress. Theoretical frameworks for cellular computation and cell-cell communication could lay the groundwork for biology to develop its first principled quantitative theories, mirroring the remarkable achievements of physics in the early 20th century. This prospect could be within our reach. What can propel us is the application of a systematic Hilbertian approach for the formalization of system-level problems in biology.

\paragraph*{Acknowledgments}
The author would like to express gratitude to the organizers and attendees of the \textit{``What is biological computation?''} workshop at \textit{Santa Fe Institute}.  Some of the ideas presented in this work were inspired by discussions that took place during the workshop. The author also wishes to thank the anonymous reviewer and Iain Johnston for their constructive comments and thoughtful suggestions.

%
%

\bibliographystyle{ieeetr}
 \bibliography{CellProblem}

\begin{thebibliography}{100}

\bibitem{Carlson2002}
J.~M. Carlson and J.~Doyle, ``Complexity and robustness,'' {\em Proceedings of
  the National Academy of Sciences}, vol.~99, no.~suppl\_1, pp.~2538--2545,
  2002.

\bibitem{Csete2002}
M.~E. Csete and J.~C. Doyle, ``Reverse engineering of biological complexity,''
  {\em Science}, vol.~295, no.~5560, pp.~1664--1669, 2002.

\bibitem{Kitano2004}
H.~Kitano, ``Biological robustness,'' {\em Nature Reviews Genetics}, vol.~5,
  pp.~826--837, Nov 2004.

\bibitem{Barkai1997}
N.~Barkai and S.~Leibler, ``Robustness in simple biochemical networks,'' {\em
  Nature}, vol.~387, pp.~913--917, Jun 1997.

\bibitem{Balazsi2011}
G.~Balazsi, A.~van Oudenaarden, and J.~J. Collins, ``Cellular decision making
  and biological noise: from microbes to mammals,'' {\em Cell}, vol.~144,
  no.~6, pp.~910--925, 2011.

\bibitem{Kitano2007}
H.~Kitano, ``Towards a theory of biological robustness,'' {\em Molecular
  Systems Biology}, vol.~3, no.~1, p.~137, 2007.

\bibitem{lazebnik2002}
Y.~Lazebnik, ``{Can a biologist fix a radio?—{Or}, what {I} learned while
  studying apoptosis},'' {\em Cancer Cell}, vol.~2, pp.~179--182, sep 2002.

\bibitem{Hilbert1902}
D.~Hilbert, ``{Mathematical problems},'' {\em Bulletin of the American
  Mathematical Society}, vol.~8, no.~10, pp.~437--479, 1902.

\bibitem{Mahner1997}
M.~Mahner and M.~Bunge, {\em Foundations of biophilosophy}.
\newblock Springer Berlin, first~ed., 1997.

\bibitem{Woodger1962}
J.~H. Woodger, ``{Biology and the Axiomatic Method},'' {\em Annals of the New
  York Academy of Sciences}, vol.~96, no.~4, pp.~1093--1116, 1962.

\bibitem{Pearl2000}
J.~Pearl, {\em Causality: models, reasoning, and inference}.
\newblock Cambridge, U.K. ; New York: Cambridge University Press, 2000.

\bibitem{Pearl2009}
J.~Pearl, ``{Causal inference in statistics: An overview},'' {\em Statistics
  Surveys}, vol.~3, no.~none, pp.~96 -- 146, 2009.

\bibitem{Weinan2011}
E.~Weinan and L.~Jianfeng, ``{M}ultiscale modeling,'' {\em Scholarpedia},
  vol.~6, no.~10, p.~11527, 2011.
\newblock revision \#91540.

\bibitem{Noble2008}
D.~Noble, ``Genes and causation,'' {\em Philosophical Transactions of the Royal
  Society A: Mathematical, Physical and Engineering Sciences}, vol.~366,
  no.~1878, pp.~3001--3015, 2008.

\bibitem{Noble2012}
D.~Noble, ``A theory of biological relativity: no privileged level of
  causation,'' {\em Interface Focus}, vol.~2, no.~1, pp.~55--64, 2012.

\bibitem{Hoel2013}
E.~P. Hoel, L.~Albantakis, and G.~Tononi, ``{Quantifying causal emergence shows
  that macro can beat micro},'' {\em Proceedings of the National Academy of
  Sciences}, vol.~110, no.~49, pp.~19790--19795, 2013.

\bibitem{Hoel2020}
E.~Hoel and M.~Levin, ``{Emergence of informative higher scales in biological
  systems: a computational toolkit for optimal prediction and control},'' {\em
  Communicative \& Integrative Biology}, vol.~13, no.~1, pp.~108--118, 2020.
\newblock PMID: 33014263.

\bibitem{Hoel2017}
E.~P. Hoel, ``When the map is better than the territory,'' {\em Entropy},
  vol.~19, no.~5, 2017.

\bibitem{Crecraft2002}
D.~Crecraft and S.~Gergely, {\em Analog Electronics: circuits, systems and
  signal processing}.
\newblock Elsevier, 2002.

\bibitem{Morris2011}
A.~S. Morris and R.~Langari, {\em Measurement and instrumentation: theory and
  application}.
\newblock Academic Press, 2011.

\bibitem{Rosenblueth1943}
A.~Rosenblueth, N.~Wiener, and J.~Bigelow, ``{Behavior, purpose and
  teleology},'' {\em Philosophy of science}, vol.~10, no.~1, pp.~18--24, 1943.

\bibitem{Ayala1970}
F.~J. Ayala, ``Teleological explanations in evolutionary biology,'' {\em
  Philosophy of Science}, vol.~37, no.~1, p.~1–15, 1970.

\bibitem{McDonough2020}
J.~K. McDonough, ``Not dead yet: Teleology and the “scientific revolution,''
  in {\em {Teleology: A History}}, p.~150–179, Oxford University Press, 07
  2020.

\bibitem{Wouters2003}
A.~G. Wouters, ``Four notions of biological function,'' {\em Studies in History
  and Philosophy of Science Part C: Studies in History and Philosophy of
  Biological and Biomedical Sciences}, vol.~34, no.~4, pp.~633--668, 2003.

\bibitem{Wimsatt19721}
W.~C. Wimsatt, ``Teleology and the logical structure of function statements,''
  {\em Studies in History and Philosophy of Science Part A}, vol.~3, no.~1,
  pp.~1--80, 1972.

\bibitem{Garcia2024}
M.~García-Valdecasas and T.~W. Deacon, ``Biological functions are causes, not
  effects: A critique of selected effects theories,'' {\em Studies in History
  and Philosophy of Science}, vol.~103, pp.~20--28, 2024.

\bibitem{Wiener2019}
N.~Wiener, {\em Cybernetics or Control and Communication in the Animal and the
  Machine}.
\newblock Cambridge, MA: MIT Press, 1965.

\bibitem{Vittadello2022}
S.~T. Vittadello and M.~P. Stumpf, ``Open problems in mathematical biology,''
  {\em Mathematical Biosciences}, vol.~354, p.~108926, 2022.

\bibitem{Selling2004}
J.~Stelling, U.~Sauer, Z.~Szallasi, F.~J. Doyle, and J.~Doyle, ``Robustness of
  cellular functions,'' {\em Cell}, vol.~118, no.~6, pp.~675--685, 2004.

\bibitem{Skocelas2022}
{\em {The Evolution of Genetic Robustness for Cellular Cooperation in Early
  Multicellular Organisms}}, vol.~ALIFE 2022: The 2022 Conference on Artificial
  Life of {\em Artificial Life Conference Proceedings}, 07 2022.

\bibitem{Alon2002}
R.~Milo, S.~Shen-Orr, S.~Itzkovitz, N.~Kashtan, D.~Chklovskii, and U.~Alon,
  ``Network motifs: Simple building blocks of complex networks,'' {\em
  Science}, vol.~298, no.~5594, pp.~824--827, 2002.

\bibitem{Alon2007}
U.~Alon, ``Network motifs: theory and experimental approaches,'' {\em Nature
  Reviews Genetics}, vol.~8, no.~6, pp.~450--461, 2007.

\bibitem{Alon2019}
U.~Alon, {\em An Introduction to Systems Biology: Design Principles of
  Biological Circuits}.
\newblock Chapman and Hall/CRC, 2019.

\bibitem{Note1}
Karl Popper highlighted the role of \protect \emph {`trial and error'} as a
  fundamental process in knowledge acquisition within evolutionary epistemology
  \cite {Popper2005}.

\bibitem{Ashby1945}
W.~R. Ashby, ``{The Physical Origin of Adaptation by Trial and Error},'' {\em
  The Journal of General Psychology}, vol.~32, no.~1, pp.~13--25, 1945.

\bibitem{Ashby2013}
W.~Ashby, {\em Design for a brain: The origin of adaptive behaviour}.
\newblock Springer Science \& Business Media, 2013.

\bibitem{Yu2006}
H.~Yu and M.~Gerstein, ``{Genomic analysis of the hierarchical structure of
  regulatory networks},'' {\em Proceedings of the National Academy of
  Sciences}, vol.~103, no.~40, pp.~14724--14731, 2006.

\bibitem{Tonner2015}
P.~D. Tonner, A.~M.~C. Pittman, J.~G. Gulli, K.~Sharma, and A.~K. Schmid, ``{A
  Regulatory Hierarchy Controls the Dynamic Transcriptional Response to Extreme
  Oxidative Stress in Archaea},'' {\em PLOS Genetics}, vol.~11, pp.~1--13, 01
  2015.

\bibitem{Defoort2018}
J.~Defoort, Y.~Van~de Peer, and V.~Vermeirssen, ``{Function, dynamics and
  evolution of network motif modules in integrated gene regulatory networks of
  worm and plant},'' {\em Nucleic Acids Research}, vol.~46, pp.~6480--6503, 06
  2018.

\bibitem{Kreimer2008}
A.~Kreimer, E.~Borenstein, U.~Gophna, and E.~Ruppin, ``The evolution of
  modularity in bacterial metabolic networks,'' {\em Proceedings of the
  National Academy of Sciences}, vol.~105, no.~19, pp.~6976--6981, 2008.

\bibitem{Kashtan2005}
N.~Kashtan and U.~Alon, ``Spontaneous evolution of modularity and network
  motifs,'' {\em Proceedings of the National Academy of Sciences}, vol.~102,
  no.~39, pp.~13773--13778, 2005.

\bibitem{Stone2019}
L.~Stone, D.~Simberloff, and Y.~Artzy-Randrup, ``Network motifs and their
  origins,'' {\em PLOS Computational Biology}, vol.~15, pp.~1--7, 04 2019.

\bibitem{Clune2013}
J.~Clune, J.-B. Mouret, and H.~Lipson, ``The evolutionary origins of
  modularity,'' {\em Proceedings of the Royal Society B: Biological Sciences},
  vol.~280, no.~1755, p.~20122863, 2013.

\bibitem{Angulo2015}
M.~T. Angulo, Y.-Y. Liu, and J.-J. Slotine, ``Network motifs emerge from
  interconnections that favour stability,'' {\em Nature Physics}, vol.~11,
  pp.~848--852, Oct 2015.

\bibitem{Bechtel2021}
W.~Bechtel and L.~Bich, ``Grounding cognition: heterarchical control mechanisms
  in biology,'' {\em Philosophical Transactions of the Royal Society B:
  Biological Sciences}, vol.~376, no.~1820, p.~20190751, 2021.

\bibitem{Simon1962}
H.~A. Simon, ``The architecture of complexity,'' {\em Proceedings of the
  American Philosophical Society}, vol.~106, no.~6, pp.~467--482, 1962.

\bibitem{McCulloch1945}
W.~S. McCulloch, ``A heterarchy of values determined by the topology of nervous
  nets,'' {\em The bulletin of mathematical biophysics}, vol.~7, pp.~89--93,
  Jun 1945.

\bibitem{Bei2013}
X.~Bei, N.~Chen, and S.~Zhang, ``{On the Complexity of Trial and Error},'' in
  {\em Proceedings of the Forty-Fifth Annual ACM Symposium on Theory of
  Computing}, STOC '13, (New York, NY, USA), p.~31–40, Association for
  Computing Machinery, 2013.

\bibitem{Note2}
A dichotomoy of view is shaping in the field; one in which, all these forms of
  internal control mechanisms are purely of evolutionary nature, and an
  opposing emergent view that adds a layer of learning and agential control to
  the single cell level control and decision making \cite {Lyon2021}.
  Interested readers may refer to a two-part special issue (in Philosophical
  Transaction B of Royal Society, 15 March 2021, Volume 376, Issue 1820) on
  this topic, entitled `Basal cognition: conceptual tools and the view from the
  single cell'.

\bibitem{Kirschner1998}
M.~Kirschner and J.~Gerhart, ``{Evolvability},'' {\em Proceedings of the
  National Academy of Sciences}, vol.~95, no.~15, pp.~8420--8427, 1998.

\bibitem{Navlakha2014}
S.~Navlakha, X.~He, C.~Faloutsos, and Z.~Bar-Joseph, ``{Topological properties
  of robust biological and computational networks},'' {\em Journal of The Royal
  Society Interface}, vol.~11, no.~96, p.~20140283, 2014.

\bibitem{Ahmed2020}
S.~S. Ahmed, S.~Roy, and J.~Kalita, ``Assessing the {Effectiveness} of
  {Causality} {Inference} {Methods} for {Gene} {Regulatory} {Networks},'' {\em
  IEEE/ACM Transactions on Computational Biology and Bioinformatics}, vol.~17,
  pp.~56--70, jan 2020.

\bibitem{Aalto2020}
A.~Aalto, L.~Viitasaari, P.~Ilmonen, L.~Mombaerts, and J.~Gon{\c{c}}alves,
  ``Gene regulatory network inference from sparsely sampled noisy data,'' {\em
  Nature Communications}, vol.~11, p.~3493, Jul 2020.

\bibitem{Lu2021}
J.~Lu, B.~Dumitrascu, I.~C. McDowell, B.~Jo, A.~Barrera, L.~K. Hong, S.~M.
  Leichter, T.~E. Reddy, and B.~E. Engelhardt, ``{Causal network inference from
  gene transcriptional time-series response to glucocorticoids},'' {\em PLOS
  Computational Biology}, vol.~17, pp.~1--29, 01 2021.

\bibitem{Shinar2006}
G.~Shinar, E.~Dekel, T.~Tlusty, and U.~Alon, ``{Rules for biological regulation
  based on error minimization},'' {\em Proceedings of the National Academy of
  Sciences}, vol.~103, no.~11, pp.~3999--4004, 2006.

\bibitem{Kang2010}
E.~Y. Kang, C.~Ye, I.~Shpitser, and E.~Eskin, ``Detecting the presence and
  absence of causal relationships between expression of yeast genes with very
  few samples,'' {\em Journal of Computational Biology}, vol.~17, no.~3,
  pp.~533--546, 2010.
\newblock PMID: 20377462.

\bibitem{Meinshausen2016}
N.~Meinshausen, A.~Hauser, J.~M. Mooij, J.~Peters, P.~Versteeg, and
  P.~Bühlmann, ``Methods for causal inference from gene perturbation
  experiments and validation,'' {\em Proceedings of the National Academy of
  Sciences}, vol.~113, no.~27, pp.~7361--7368, 2016.

\bibitem{Wen2023}
Y.~Wen, J.~Huang, S.~Guo, Y.~Elyahu, A.~Monsonego, H.~Zhang, Y.~Ding, and
  H.~Zhu, ``Applying causal discovery to single-cell analyses using
  causalcell,'' {\em eLife}, vol.~12, p.~e81464, may 2023.

\bibitem{Lotito2022}
Q.~F. Lotito, F.~Musciotto, A.~Montresor, and F.~Battiston, ``Higher-order
  motif analysis in hypergraphs,'' {\em Communications Physics}, vol.~5, p.~79,
  Apr 2022.

\bibitem{Battiston2021}
F.~Battiston, E.~Amico, A.~Barrat, G.~Bianconi, G.~Ferraz~de Arruda,
  B.~Franceschiello, I.~Iacopini, S.~K{\'e}fi, V.~Latora, Y.~Moreno, M.~M.
  Murray, T.~P. Peixoto, F.~Vaccarino, and G.~Petri, ``The physics of
  higher-order interactions in complex systems,'' {\em Nature Physics},
  vol.~17, pp.~1093--1098, Oct 2021.

\bibitem{Tabar2024}
M.~R.~R. Tabar, F.~Nikakhtar, L.~Parkavousi, A.~Akhshi, U.~Feudel, and
  K.~Lehnertz, ``Revealing higher-order interactions in high-dimensional
  complex systems: A data-driven approach,'' {\em Phys. Rev. X}, vol.~14,
  p.~011050, Mar 2024.

\bibitem{Note3}
This approach is similar to how in-circuit testing for printed circuit boards
  works, where the testing is not just about understanding the connections, but
  also about probing the functionality of the individual components and the
  system as a whole. This allows for a more comprehensive understanding of the
  system, whether it’s a circuit board or a biological cell.

\bibitem{Lamport1978}
L.~Lamport, ``{Time, Clocks and the Ordering of Events in a Distributed
  System},'' {\em Communications of the ACM}, vol.~21, pp.~558--565, July 1978.

\bibitem{Regev2001}
A.~Regev, W.~Silverman, and E.~Shapiro, ``{Representation and simulation of
  biochemical processes using the pi-calculus process algebra},'' {\em Pac Symp
  Biocomput}, vol.~1, pp.~459--470, 2001.

\bibitem{Priami2001}
C.~Priami, A.~Regev, E.~Shapiro, and W.~Silverman, ``{Application of a
  stochastic name-passing calculus to representation and simulation of
  molecular processes},'' {\em Information Processing Letters}, vol.~80, no.~1,
  pp.~25--31, 2001.
\newblock Process Algebra.

\bibitem{Pinney2003}
J.~Pinney, D.~Westhead, and G.~McConkey, ``{Petri Net representations in
  systems biology},'' {\em Biochemical Society Transactions}, vol.~31,
  pp.~1513--1515, 12 2003.

\bibitem{Phillips2007}
A.~Phillips and L.~Cardelli, ``{Efficient, Correct Simulation of Biological
  Processes in the Stochastic Pi-calculus},'' {\em Computational Methods in
  Systems Biology}, vol.~4695, pp.~184--199, September 2007.

\bibitem{Priami2009}
C.~Priami, ``{Algorithmic Systems Biology},'' {\em Commun. ACM}, vol.~52,
  p.~80–88, may 2009.

\bibitem{Cooper2008}
S.~J. Cooper, ``{From Claude Bernard to Walter Cannon. Emergence of the concept
  of homeostasis},'' {\em Appetite}, vol.~51, no.~3, pp.~419--427, 2008.

\bibitem{Cannon1929}
W.~B. Cannon, ``{Organization for physiological homeostasis},'' {\em
  Physiological Reviews}, vol.~9, no.~3, pp.~399--431, 1929.

\bibitem{Modell2015}
H.~Modell, W.~Cliff, J.~Michael, J.~McFarland, M.~P. Wenderoth, and A.~Wright,
  ``{A physiologist's view of homeostasis},'' {\em Advances in Physiology
  Education}, vol.~39, no.~4, pp.~259--266, 2015.
\newblock PMID: 26628646.

\bibitem{Schneck1987}
D.~Schneck, ``{Feedback control and the concept of homeostasis},'' {\em
  Mathematical Modelling}, vol.~9, no.~12, pp.~889--900, 1987.

\bibitem{Yi2000}
T.-M. Yi, Y.~Huang, M.~I. Simon, and J.~Doyle, ``{Robust perfect adaptation in
  bacterial chemotaxis through integral feedback control},'' {\em Proceedings
  of the National Academy of Sciences}, vol.~97, no.~9, pp.~4649--4653, 2000.

\bibitem{Briat2016}
C.~Briat, A.~Gupta, and M.~Khammash, ``{Antithetic Integral Feedback Ensures
  Robust Perfect Adaptation in Noisy Biomolecular Networks},'' {\em Cell
  Systems}, vol.~2, no.~1, pp.~15--26, 2016.

\bibitem{Aoki2019}
S.~K. Aoki, G.~Lillacci, A.~Gupta, A.~Baumschlager, D.~Schweingruber, and
  M.~Khammash, ``A universal biomolecular integral feedback controller for
  robust perfect adaptation,'' {\em Nature}, vol.~570, pp.~533--537, jun 2019.

\bibitem{Chevalier2019}
M.~Chevalier, M.~G{\'o}mez-Schiavon, A.~H. Ng, and H.~El-Samad, ``Design and
  analysis of a proportional-integral-derivative controller with biological
  molecules,'' {\em Cell Systems}, vol.~9, pp.~338--353.e10, Oct 2019.

\bibitem{Qian2022}
Y.~Qian, D.~Del~Vecchio, and H.~Qian, ``Antithetic integral feedback enables
  robust adaptation in stochastic biomolecular networks,'' {\em Nature
  Communications}, vol.~13, no.~1, pp.~1--14, 2022.

\bibitem{Filo2022}
M.~Filo, S.~Kumar, and M.~Khammash, ``A hierarchy of biomolecular
  proportional-integral-derivative feedback controllers for robust perfect
  adaptation and dynamic performance,'' {\em Nature Communications}, vol.~13,
  p.~2119, Apr 2022.

\bibitem{Note4}
May be referred to as \protect \emph {Allostasis} \cite {McEwen1998,
  McEwen1998b}, \protect \emph {Heterostasis} \cite {Selye1973}, or \protect
  \emph {Adaptive Homeostasis} \cite {Davies2016}.

\bibitem{Waddington1957}
C.~Waddington, {\em The Strategy of the Genes}.
\newblock Routledge, 1st~ed., 1957.

\bibitem{Note5}
The Waddington landscape model offers a multidimensional perspective on
  cellular behavior, illustrating various potential states of cell development
  as attractor basins within a complex topography \cite {Waddington1957}. This
  model highlights the role of noise in shaping the epigenetic landscape, as
  fluctuations can distort the pathways that cells follow, leading to diverse
  cell-fate decisions \cite {Coomer2022}. Noise, therefore, is not merely a
  disruptive force but a crucial element that can drive variability and
  adaptability in biological systems. Modern interpretations of this model show
  how bifurcations and bistability are necessary for understanding the
  robustness and flexibility of cellular decisions and developmental pathways
  \cite {Ferrell2012}. The intrinsic noisiness of gene networks is another
  aspect where noise plays a pivotal role. The fidelity of molecular signaling
  is influenced by this inherent noise, impacting the information capacity and
  reliability of these regulatory systems \cite {Paulsson2004, Paulsson2005}.
  These perspectives collectively emphasize the importance of noise in both
  genetic regulation and epigenetic landscapes, illustrating how stochasticity
  can be harnessed for functional adaptability and complexity in cellular
  systems.

\bibitem{Aderem2005}
A.~Aderem, ``Systems biology: Its practice and challenges,'' {\em Cell},
  vol.~121, pp.~511--513, May 2005.

\bibitem{Klein2020}
B.~Klein and E.~Hoel, ``The emergence of informative higher scales in complex
  networks,'' {\em Complexity}, vol.~2020, no.~1, p.~8932526, 2020.

\bibitem{Klein2021}
B.~Klein, E.~Hoel, A.~Swain, R.~Griebenow, and M.~Levin, ``{Evolution and
  emergence: higher order information structure in protein interactomes across
  the tree of life},'' {\em Integrative Biology}, vol.~13, pp.~283--294, 12
  2021.

\bibitem{Sterling2012}
P.~Sterling, ``{Allostasis: A model of predictive regulation},'' {\em
  Physiology \& Behavior}, vol.~106, no.~1, pp.~5--15, 2012.
\newblock Allostasis and Allostatic Load.

\bibitem{Franciolini1989}
F.~Franciolini and A.~Petris, ``{Evolution of ionic channels of biological
  membranes.},'' {\em Molecular Biology and Evolution}, vol.~6, pp.~503--513,
  09 1989.

\bibitem{Moran2015}
Y.~Moran, M.~G. Barzilai, B.~J. Liebeskind, H.~H. Zakon, and P.~A.~V. Anderson,
  ``{Evolution of voltage-gated ion channels at the emergence of Metazoa},''
  {\em Journal of Experimental Biology}, vol.~218, pp.~515--525, 02 2015.

\bibitem{Wan2021}
K.~Y. Wan and G.~Jékely, ``Origins of eukaryotic excitability,'' {\em
  Philosophical Transactions of the Royal Society B: Biological Sciences},
  vol.~376, no.~1820, p.~20190758, 2021.

\bibitem{Oleary2018}
T.~O’Leary, ``Homeostasis, failure of homeostasis and degenerate ion channel
  regulation,'' {\em Current Opinion in Physiology}, vol.~2, pp.~129--138,
  2018.
\newblock Ion Channels.

\bibitem{Williams2013}
A.~H. Williams, T.~O'Leary, and E.~Marder, ``{H}omeostatic {R}egulation of
  {N}euronal {E}xcitability,'' {\em Scholarpedia}, vol.~8, no.~1, p.~1656,
  2013.
\newblock revision \#140978.

\bibitem{Schauf1987}
C.~L. Schauf, ``Ion channel diversity: a revolution in biology?,'' {\em Science
  Progress (1933-)}, pp.~459--478, 1987.

\bibitem{Anderson2017}
P.~A. Anderson and R.~M. Greenberg, ``Phylogeny of ion channels: clues to
  structure and function,'' {\em Comparative Biochemistry and Physiology Part
  B: Biochemistry and Molecular Biology}, vol.~129, no.~1, pp.~17--28, 2001.

\bibitem{Doyle1998}
D.~A. Doyle, J.~M. Cabral, R.~A. Pfuetzner, A.~Kuo, J.~M. Gulbis, S.~L. Cohen,
  B.~T. Chait, and R.~MacKinnon, ``{The Structure of the Potassium Channel:
  Molecular Basis of K+ Conduction and Selectivity},'' {\em Science}, vol.~280,
  no.~5360, pp.~69--77, 1998.

\bibitem{Linder2013}
T.~Linder, B.~L. de~Groot, and A.~Stary-Weinzinger, ``{Probing the Energy
  Landscape of Activation Gating of the Bacterial Potassium Channel KcsA},''
  {\em PLOS Computational Biology}, vol.~9, pp.~1--9, 05 2013.

\bibitem{Hirano2011}
M.~Hirano, Y.~Onishi, T.~Yanagida, and T.~Ide, ``{Role of the KcsA channel
  cytoplasmic domain in pH-dependent gating},'' {\em Biophysical journal},
  vol.~101, no.~9, pp.~2157--2162, 2011.

\bibitem{Baker2007}
K.~A. Baker, C.~Tzitzilonis, W.~Kwiatkowski, S.~Choe, and R.~Riek,
  ``{Conformational dynamics of the {KcsA} potassium channel governs gating
  properties},'' {\em Nature Structural \& Molecular Biology}, vol.~14,
  pp.~1089--1095, nov 2007.

\bibitem{Gao2005}
L.~Gao, X.~Mi, V.~Paajanen, K.~Wang, and Z.~Fan, ``{Activation-coupled
  inactivation in the bacterial potassium channel KcsA},'' {\em Proceedings of
  the National Academy of Sciences}, vol.~102, no.~49, pp.~17630--17635, 2005.

\bibitem{Booth2007}
I.~R. Booth, {\em The Regulation of Intracellular pH in Bacteria}, ch.~3,
  pp.~19--37.
\newblock John Wiley $\&$ Sons, Ltd, 2007.

\bibitem{Putnam2012}
R.~W. Putnam, ``{Chapter 17 - Intracellular pH Regulation},'' in {\em Cell
  Physiology Source Book (Fourth Edition)} (N.~Sperelakis, ed.), pp.~303--321,
  San Diego: Academic Press, fourth edition~ed., 2012.

\bibitem{Isfort1993}
R.~J. Isfort, D.~B. Cody, T.~N. Asquith, G.~M. Ridder, S.~B. Stuard, and R.~A.
  Leboeuf, ``{Induction of protein phosphorylation, protein synthesis,
  immediate-early-gene expression and cellular proliferation by intracellular
  pH modulation},'' {\em European Journal of Biochemistry}, vol.~213, no.~1,
  pp.~349--357, 1993.

\bibitem{Note6}
Detailed reviews of mathematical models and simulations of ion channels can be
  found in \cite {Roux2004, Southern2008, Maffeo2012, Guardiani2022,
  Clerx2019}. These works explore various approaches, including traditional
  methods like Hodgkin-Huxley and Markov models, as well as modern techniques
  such as Artificial Neural Networks (ANNs) and Neural Ordinary Differential
  Equations (ODEs) for modeling ion channel conductance and kinetics \cite
  {Jeong2021, Lei2021}. These advanced models offer improved accuracy and
  computational performance, providing valuable tools for studying the
  electrophysiological mechanisms of ion channels across different biological
  systems.

\bibitem{Langthaler2022}
S.~Langthaler, J.~Lozanović~Šajić, T.~Rienmüller, S.~H. Weinberg, and
  C.~Baumgartner, ``Ion channel modeling beyond state of the art: A comparison
  with a system theory-based model of the shaker-related voltage-gated
  potassium channel kv1.1,'' {\em Cells}, vol.~11, no.~2, 2022.

\bibitem{Oleary2011}
T.~O’Leary and D.~J.~A. Wyllie, ``Neuronal homeostasis: time for a change?,''
  {\em The Journal of Physiology}, vol.~589, no.~20, pp.~4811--4826, 2011.

\bibitem{Dehghani2023}
N.~Dehghani and G.~Caterina, ``Physical computing: A category theoretic
  perspective on physical computation and system compositionality,'' 2023.

\bibitem{Oleary2013}
T.~O{\textquoteright}Leary, A.~H. Williams, J.~S. Caplan, and E.~Marder,
  ``{Correlations in ion channel expression emerge from homeostatic tuning
  rules},'' {\em Proceedings of the National Academy of Sciences}, vol.~110,
  no.~28, pp.~E2645--E2654, 2013.

\bibitem{Ori2018}
H.~Ori, E.~Marder, and S.~Marom, ``{Cellular function given parametric
  variation in the Hodgkin and Huxley model of excitability},'' {\em
  Proceedings of the National Academy of Sciences}, vol.~115, no.~35,
  pp.~E8211--E8218, 2018.

\bibitem{Yang2022}
J.~Yang, H.~Shakil, S.~Ratté, and S.~A. Prescott, ``Minimal requirements for a
  neuron to coregulate many properties and the implications for ion channel
  correlations and robustness,'' {\em eLife}, vol.~11, p.~e72875, mar 2022.

\bibitem{Dumenieu2017}
M.~Duménieu, M.~Oulé, M.~R. Kreutz, and J.~Lopez-Rojas, ``{The Segregated
  Expression of Voltage-Gated Potassium and Sodium Channels in Neuronal
  Membranes: Functional Implications and Regulatory Mechanisms},'' {\em
  Frontiers in Cellular Neuroscience}, vol.~11, p.~115, 2017.

\bibitem{Meadows2015}
J.~P. Meadows, M.~C. Guzman-Karlsson, S.~Phillips, C.~Holleman, J.~L. Posey,
  J.~J. Day, J.~J. Hablitz, and J.~D. Sweatt, ``{DNA methylation regulates
  neuronal glutamatergic synaptic scaling},'' {\em Science Signaling}, vol.~8,
  no.~382, pp.~ra61--ra61, 2015.

\bibitem{Meadows2016}
J.~P. Meadows, M.~C. Guzman-Karlsson, S.~Phillips, J.~A. Brown, S.~K. Strange,
  J.~D. Sweatt, and J.~J. Hablitz, ``{Dynamic DNA methylation regulates
  neuronal intrinsic membrane excitability},'' {\em Science Signaling}, vol.~9,
  no.~442, pp.~ra83--ra83, 2016.

\bibitem{Rosati2004}
B.~Rosati and D.~McKinnon, ``{Regulation of Ion Channel Expression},'' {\em
  Circulation Research}, vol.~94, no.~7, pp.~874--883, 2004.

\bibitem{Yue1999}
L.~Yue, P.~Melnyk, R.~Gaspo, Z.~Wang, and S.~Nattel, ``{Molecular Mechanisms
  Underlying Ionic Remodeling in a Dog Model of Atrial Fibrillation},'' {\em
  Circulation Research}, vol.~84, no.~7, pp.~776--784, 1999.

\bibitem{Balse2012}
E.~Balse, D.~F. Steele, H.~Abriel, A.~Coulombe, D.~Fedida, and S.~N. Hatem,
  ``{Dynamic of Ion Channel Expression at the Plasma Membrane of
  Cardiomyocytes},'' {\em Physiological Reviews}, vol.~92, no.~3,
  pp.~1317--1358, 2012.
\newblock PMID: 22811429.

\bibitem{Tsimring2014}
L.~S. Tsimring, ``{Noise in biology},'' {\em Reports on Progress in Physics},
  vol.~77, p.~026601, jan 2014.

\bibitem{Azpeitia2020}
E.~Azpeitia, E.~P. Balanzario, and A.~Wagner, ``{Signaling pathways have an
  inherent need for noise to acquire information},'' {\em BMC Bioinformatics},
  vol.~21, p.~462, Oct 2020.

\bibitem{Newman1990}
A.~H. Newman, ``{Chapter 29. Irreversible Ligands for Drug Receptor
  Characterization},'' in {\em Annual Reports in Medicinal Chemistry} (J.~A.
  Bristol, ed.), vol.~25, pp.~271--280, Academic Press, 1990.

\bibitem{Pierobon2011}
M.~Pierobon and I.~F. Akyildiz, ``{Noise Analysis in Ligand-Binding Reception
  for Molecular Communication in Nanonetworks},'' {\em IEEE Transactions on
  Signal Processing}, vol.~59, no.~9, pp.~4168--4182, 2011.

\bibitem{Lestas2010}
I.~Lestas, G.~Vinnicombe, and J.~Paulsson, ``Fundamental limits on the
  suppression of molecular fluctuations,'' {\em Nature}, vol.~467,
  pp.~174--178, Sep 2010.

\bibitem{Teimouri2020}
H.~Teimouri and A.~B. Kolomeisky, ``Relaxation times of ligand-receptor complex
  formation control t cell activation,'' {\em Biophysical Journal}, vol.~119,
  pp.~182--189, Jul 2020.

\bibitem{Duke1999}
T.~A.~J. Duke and D.~Bray, ``Heightened sensitivity of a lattice of membrane
  receptors,'' {\em Proceedings of the National Academy of Sciences}, vol.~96,
  no.~18, pp.~10104--10108, 1999.

\bibitem{Rappel2008a}
W.-J. Rappel and H.~Levine, ``Receptor noise and directional sensing in
  eukaryotic chemotaxis,'' {\em Phys. Rev. Lett.}, vol.~100, p.~228101, Jun
  2008.

\bibitem{Rappel2008b}
W.-J. Rappel and H.~Levine, ``Receptor noise limitations on chemotactic
  sensing,'' {\em Proceedings of the National Academy of Sciences}, vol.~105,
  no.~49, pp.~19270--19275, 2008.

\bibitem{Ventura2014}
A.~C. Ventura, A.~Bush, G.~Vasen, M.~A. Gold{\'\i}n, B.~Burkinshaw,
  N.~Bhattacharjee, A.~Folch, R.~Brent, A.~Chernomoretz, and A.~Colman-Lerner,
  ``{Utilization of extracellular information before ligand-receptor binding
  reaches equilibrium expands and shifts the input dynamic range},'' {\em
  Proceedings of the National Academy of Sciences}, vol.~111, no.~37,
  pp.~E3860--E3869, 2014.

\bibitem{Kajita2020}
M.~K. Kajita, K.~Aihara, and T.~J. Kobayashi, ``{Reliable target ligand
  detection by noise-induced receptor cluster formation},'' {\em Chaos: An
  Interdisciplinary Journal of Nonlinear Science}, vol.~30, p.~011104, 01 2020.

\bibitem{Aquino2011}
G.~Aquino, D.~Clausznitzer, S.~Tollis, and R.~G. Endres, ``Optimal
  receptor-cluster size determined by intrinsic and extrinsic noise,'' {\em
  Phys. Rev. E}, vol.~83, p.~021914, Feb 2011.

\bibitem{Care2013}
B.~R. Car\'e and H.~A. Soula, ``Receptor clustering affects signal transduction
  at the membrane level in the reaction-limited regime,'' {\em Phys. Rev. E},
  vol.~87, p.~012720, Jan 2013.

\bibitem{Note7}
Stochastic resonance, where noise actually enhances the detection of weak
  signals, is further elaborated in the following sections.

\bibitem{Bene2022}
L.~Bene, M.~Bagdány, and L.~Damjanovich, ``T-cell receptor is a threshold
  detector: Sub- and supra-threshold stochastic resonance in tcr-mhc clusters
  on the cell surface,'' {\em Entropy}, vol.~24, no.~3, 2022.

\bibitem{Nagano2020}
S.~Nagano, ``{Noise reduction and signal enhancement by receptor
  synchronization},'' {\em Nonlinear Theory and Its Applications, IEICE},
  vol.~11, no.~4, pp.~601--609, 2020.

\bibitem{Monod1974}
J.~Monod, ``{On Chance and Necessity},'' in {\em Studies in the Philosophy of
  Biology: Reduction and Related Problems} (F.~J. Ayala and T.~Dobzhansky,
  eds.), pp.~357--375, London: Macmillan Education UK, 1974.

\bibitem{Noble2021}
D.~Noble, ``{The role of stochasticity in biological communication
  processes},'' {\em Progress in Biophysics and Molecular Biology}, vol.~162,
  pp.~122--128, 2021.
\newblock On the Physics of Excitable Media. Waves in Soft and Living Matter,
  their Transmission at the Synapse and their Cooperation in the Brain.

\bibitem{Hille2001}
B.~Hille, {\em Ion channels of excitable membranes}.
\newblock Sunderland, Mass: Sinauer, 3rd ed~ed., 2001.

\bibitem{Sakmann1995}
B.~Sakmann and E.~Neher, {\em Single-channel recording}.
\newblock Boston, MA: Springer US, 1995.

\bibitem{Horikawa1991}
Y.~Horikawa, ``{Noise effects on spike propagation in the stochastic
  Hodgkin-Huxley models},'' {\em Biological Cybernetics}, vol.~66, pp.~19--25,
  Nov 1991.

\bibitem{White2000}
J.~A. White, J.~T. Rubinstein, and A.~R. Kay, ``{Channel noise in neurons},''
  {\em Trends in Neurosciences}, vol.~23, no.~3, pp.~131--137, 2000.

\bibitem{Schneidman1998}
E.~Schneidman, B.~Freedman, and I.~Segev, ``{Ion Channel Stochasticity May Be
  Critical in Determining the Reliability and Precision of Spike Timing},''
  {\em Neural Computation}, vol.~10, pp.~1679--1703, 10 1998.

\bibitem{Faisal2005}
A.~A. Faisal, J.~A. White, and S.~B. Laughlin, ``{Ion-Channel Noise Places
  Limits on the Miniaturization of the Brain\'s Wiring},'' {\em Current
  Biology}, vol.~15, pp.~1143--1149, Jun 2005.

\bibitem{Wiesenfeld1995}
K.~Wiesenfeld and F.~Moss, ``{Stochastic resonance and the benefits of noise:
  from ice ages to crayfish and SQUIDs},'' {\em Nature}, vol.~373, pp.~33--36,
  Jan 1995.

\bibitem{Note8}
SR was first introduced as an explanation for the observed periodic occurrences
  of the Earth's ice ages \cite {Benzi1981}. In SR, adding noise to a nonlinear
  dynamical system can bring a weak signal above the threshold, enabling the
  system to detect sub-threshold signals \cite {Nicolis2007,Gammaitoni1998}. In
  this case, the system that is subject to periodic forcing shows a resonance
  in the spectrum that is absent in the forcing and the perturbation \cite
  {Benzi1981}.

\bibitem{Zeng2000}
F.-G. Zeng, Q.-J. Fu, and R.~Morse, ``{Human hearing enhanced by noise},'' {\em
  Brain Research}, vol.~869, no.~1, pp.~251--255, 2000.

\bibitem{Simonotto1997}
E.~Simonotto, M.~Riani, C.~Seife, M.~Roberts, J.~Twitty, and F.~Moss, ``{Visual
  Perception of Stochastic Resonance},'' {\em Phys. Rev. Lett.}, vol.~78,
  pp.~1186--1189, Feb 1997.

\bibitem{Priplata2002}
A.~Priplata, J.~Niemi, M.~Salen, J.~Harry, L.~A. Lipsitz, and J.~J. Collins,
  ``{Noise-Enhanced Human Balance Control},'' {\em Phys. Rev. Lett.}, vol.~89,
  p.~238101, Nov 2002.

\bibitem{Collins1996}
J.~J. Collins, T.~T. Imhoff, and P.~Grigg, ``{Noise-enhanced tactile
  sensation},'' {\em Nature}, vol.~383, pp.~770--770, Oct 1996.

\bibitem{Collins1995}
J.~J. Collins, C.~C. Chow, and T.~T. Imhoff, ``{Stochastic resonance without
  tuning},'' {\em Nature}, vol.~376, pp.~236--238, Jul 1995.

\bibitem{Douglass1993}
J.~K. Douglass, L.~Wilkens, E.~Pantazelou, and F.~Moss, ``{Noise enhancement of
  information transfer in crayfish mechanoreceptors by stochastic resonance},''
  {\em Nature}, vol.~365, pp.~337--340, Sep 1993.

\bibitem{Bulsara1991}
A.~Bulsara, E.~Jacobs, T.~Zhou, F.~Moss, and L.~Kiss, ``{Stochastic resonance
  in a single neuron model: Theory and analog simulation},'' {\em Journal of
  Theoretical Biology}, vol.~152, no.~4, pp.~531--555, 1991.

\bibitem{Wiesenfeld1994}
K.~Wiesenfeld, D.~Pierson, E.~Pantazelou, C.~Dames, and F.~Moss, ``{Stochastic
  resonance on a circle},'' {\em Phys. Rev. Lett.}, vol.~72, pp.~2125--2129,
  Apr 1994.

\bibitem{Bezrukov1995}
S.~M. Bezrukov and I.~Vodyanoy, ``{Noise-induced enhancement of signal
  transduction across voltage-dependent ion channels},'' {\em Nature},
  vol.~378, pp.~362--364, Nov 1995.

\bibitem{Bezrukov1998}
S.~M. Bezrukov and I.~Vodyanoy, ``{Stochastic resonance in thermally activated
  reactions: Application to biological ion channels},'' {\em Chaos: An
  Interdisciplinary Journal of Nonlinear Science}, vol.~8, no.~3, pp.~557--566,
  1998.

\bibitem{Bezrukov1997}
S.~M. Bezrukov and I.~Vodyanoy, ``{Stochastic resonance in non-dynamical
  systems without response thresholds},'' {\em Nature}, vol.~385, pp.~319--321,
  Jan 1997.

\bibitem{Vincent2021}
U.~E. Vincent, P.~V.~E. McClintock, I.~A. Khovanov, and S.~Rajasekar,
  ``Vibrational and stochastic resonances in driven nonlinear systems,'' {\em
  Philosophical Transactions of the Royal Society A: Mathematical, Physical and
  Engineering Sciences}, vol.~379, no.~2192, p.~20200226, 2021.

\bibitem{Sorokin2021}
V.~Sorokin and I.~Demidov, ``On representing noise by deterministic excitations
  for interpreting the stochastic resonance phenomenon,'' {\em Philosophical
  Transactions of the Royal Society A: Mathematical, Physical and Engineering
  Sciences}, vol.~379, no.~2192, p.~20200229, 2021.

\bibitem{Lucarini2019}
V.~Lucarini, ``Stochastic resonance for nonequilibrium systems,'' {\em Phys.
  Rev. E}, vol.~100, p.~062124, Dec 2019.

\bibitem{Schmid2001}
G.~Schmid, I.~Goychuk, and P.~Hänggi, ``{Stochastic resonance as a collective
  property of ion channel assemblies},'' {\em Europhysics Letters ({EPL})},
  vol.~56, pp.~22--28, oct 2001.

\bibitem{Parc2009}
Y.~W. Parc, D.-S. Koh, and W.~Sung, ``{Stochastic resonance in an ion channel
  following the non-Arrhenius gating rate},'' {\em The European Physical
  Journal B}, vol.~69, pp.~127--131, May 2009.

\bibitem{Goychuk2000}
I.~Goychuk and P.~H\"anggi, ``{Stochastic resonance in ion channels
  characterized by information theory},'' {\em Phys. Rev. E}, vol.~61,
  pp.~4272--4280, Apr 2000.

\bibitem{Adair2003}
R.~K. Adair, ``Noise and stochastic resonance in voltage-gated ion channels,''
  {\em Proceedings of the National Academy of Sciences}, vol.~100, no.~21,
  pp.~12099--12104, 2003.

\bibitem{Toral2003}
R.~Toral, C.~R. Mirasso, and J.~D. Gunton, ``{System size coherence resonance
  in coupled {FitzHugh}-Nagumo models},'' {\em Europhysics Letters ({EPL})},
  vol.~61, pp.~162--167, jan 2003.

\bibitem{Pikovsky2002}
A.~Pikovsky, A.~Zaikin, and M.~A. de~la Casa, ``{System Size Resonance in
  Coupled Noisy Systems and in the Ising Model},'' {\em Phys. Rev. Lett.},
  vol.~88, p.~050601, Jan 2002.

\bibitem{Schmid2004}
G.~Schmid, I.~Goychuk, P.~Hanggi, S.~Zeng, and P.~Jung, ``{Stochastic resonance
  and optimal clustering for assemblies of ion channels},'' {\em Fluctuation
  and Noise Letters}, vol.~04, no.~01, pp.~L33--L42, 2004.

\bibitem{Wannamaker2000}
R.~A. Wannamaker, S.~P. Lipshitz, and J.~Vanderkooy, ``{Stochastic resonance as
  dithering},'' {\em Phys. Rev. E}, vol.~61, pp.~233--236, Jan 2000.

\bibitem{Note9}
During World War II, engineers discovered that mechanical computers used for
  radars and trajectory calculations performed better onboard airplanes than on
  the ground. They found that vibration-induced noise randomized the
  quantization error, thereby increasing the accuracy of the calculations
  performed by these mechanical computers \cite {Pohlmann1989}. Since then, it
  has become standard practice to add a small amount of noise to a signal
  either before quantization (during analog-to-digital conversion) or when
  reducing the bit-depth of a digital signal (such as during a 256-bit to a
  16-bit image reduction).

\bibitem{Roberts1962}
L.~Roberts, ``{Picture coding using pseudo-random noise},'' {\em IRE
  Transactions on Information Theory}, vol.~8, no.~2, pp.~145--154, 1962.

\bibitem{Gammaitoni1995_a}
L.~Gammaitoni, ``{Stochastic resonance and the dithering effect in threshold
  physical systems},'' {\em Phys. Rev. E}, vol.~52, pp.~4691--4698, Nov 1995.

\bibitem{Gammaitoni1995_b}
L.~Gammaitoni, ``{Stochastic resonance in multi-threshold systems},'' {\em
  Physics Letters A}, vol.~208, no.~4, pp.~315--322, 1995.

\bibitem{Hoekstra2014}
A.~Hoekstra, B.~Chopard, and P.~Coveney, ``Multiscale modelling and simulation:
  a position paper,'' {\em Philosophical Transactions of the Royal Society A:
  Mathematical, Physical and Engineering Sciences}, vol.~372, no.~2021,
  p.~20130377, 2014.

\bibitem{Laughlin2000}
R.~B. Laughlin, D.~Pines, J.~Schmalian, B.~P. Stojković, and P.~Wolynes, ``The
  middle way,'' {\em Proceedings of the National Academy of Sciences}, vol.~97,
  no.~1, pp.~32--37, 2000.

\bibitem{Rice2024}
C.~Rice, ``Beyond reduction and emergence: a framework for tailoring multiscale
  modeling techniques to specific contexts,'' {\em Biology {\&} Philosophy},
  vol.~39, p.~12, Jun 2024.

\bibitem{Weinan2003}
W.~E, B.~Engquist, and Z.~Huang, ``Heterogeneous multiscale method: A general
  methodology for multiscale modeling,'' {\em Phys. Rev. B}, vol.~67,
  p.~092101, Mar 2003.

\bibitem{Noble2018}
R.~Noble and D.~Noble, ``Harnessing stochasticity: How do organisms make
  choices?,'' {\em Chaos: An Interdisciplinary Journal of Nonlinear Science},
  vol.~28, no.~10, p.~106309, 2018.

\bibitem{Gunawardena2014}
J.~Gunawardena, ``{Models in biology: `accurate descriptions of our pathetic
  thinking'},'' {\em BMC Biology}, vol.~12, p.~29, Apr 2014.

\bibitem{Bialek2012}
W.~S. Bialek, {\em Biophysics: searching for principles}.
\newblock Princeton, NJ: Princeton University Press, 2012.

\bibitem{Navlakha2011}
S.~Navlakha and Z.~Bar-Joseph, ``{Algorithms in nature: the convergence of
  systems biology and computational thinking},'' {\em Molecular Systems
  Biology}, vol.~7, no.~1, p.~546, 2011.

\bibitem{Krakauer2011}
D.~C. Krakauer, J.~P. Collins, D.~Erwin, J.~C. Flack, W.~Fontana, M.~D.
  Laubichler, S.~J. Prohaska, G.~B. West, and P.~F. Stadler, ``The challenges
  and scope of theoretical biology,'' {\em Journal of Theoretical Biology},
  vol.~276, no.~1, pp.~269--276, 2011.

\bibitem{Mayr2004}
E.~Mayr, {\em What makes biology unique?: considerations on the autonomy of a
  scientific discipline}.
\newblock New York: Cambridge University Press, 2004.

\bibitem{Rackauckas2020}
C.~Rackauckas, Y.~Ma, J.~Martensen, C.~Warner, K.~Zubov, R.~Supekar,
  D.~Skinner, A.~Ramadhan, and A.~Edelman, ``{Universal Differential Equations
  for Scientific Machine Learning},'' {\em arXiv}, vol.~1, 2020.

\bibitem{Karniadakis2021}
G.~E. Karniadakis, I.~G. Kevrekidis, L.~Lu, P.~Perdikaris, S.~Wang, and
  L.~Yang, ``Physics-informed machine learning,'' {\em Nature Reviews Physics},
  vol.~3, pp.~422--440, Jun 2021.

\bibitem{Lu2022}
P.~Y. Lu, J.~Ari{\~{n}}o~Bernad, and M.~Solja{\v{c}}i{\'{c}}, ``Discovering
  sparse interpretable dynamics from partial observations,'' {\em
  Communications Physics}, vol.~5, p.~206, Aug 2022.

\bibitem{Reinbold2021}
P.~A.~K. Reinbold, L.~M. Kageorge, M.~F. Schatz, and R.~O. Grigoriev, ``Robust
  learning from noisy, incomplete, high-dimensional experimental data via
  physically constrained symbolic regression,'' {\em Nature Communications},
  vol.~12, p.~3219, May 2021.

\bibitem{Groen2019}
D.~Groen, J.~Knap, P.~Neumann, D.~Suleimenova, L.~Veen, and K.~Leiter,
  ``Mastering the scales: a survey on the benefits of multiscale computing
  software,'' {\em Philosophical Transactions of the Royal Society A:
  Mathematical, Physical and Engineering Sciences}, vol.~377, no.~2142,
  p.~20180147, 2019.

\bibitem{Alber2019}
M.~Alber, A.~Buganza~Tepole, W.~R. Cannon, S.~De, S.~Dura-Bernal,
  K.~Garikipati, G.~Karniadakis, W.~W. Lytton, P.~Perdikaris, L.~Petzold, and
  E.~Kuhl, ``Integrating machine learning and multiscale
  modeling---perspectives, challenges, and opportunities in the biological,
  biomedical, and behavioral sciences,'' {\em npj Digital Medicine}, vol.~2,
  p.~115, Nov 2019.

\bibitem{Vlachas2022}
P.~R. Vlachas, G.~Arampatzis, C.~Uhler, and P.~Koumoutsakos, ``Multiscale
  simulations of complex systems by learning their effective dynamics,'' {\em
  Nature Machine Intelligence}, vol.~4, pp.~359--366, Apr 2022.

\end{thebibliography}





\end{document}